\documentclass[pdflatex,sn-mathphys-num]{sn-jnl}

\usepackage{graphicx}%
\usepackage{multirow}%
\usepackage{amsmath,amssymb,amsfonts}%
\usepackage{amsthm}%
\usepackage{mathrsfs}%
\usepackage[title]{appendix}%
\usepackage{xcolor}%
\usepackage{textcomp}%
\usepackage{manyfoot}%
\usepackage{booktabs}%
\usepackage{algorithm}%
\usepackage{algorithmicx}%
\usepackage{algpseudocode}%
\usepackage{listings}%
\usepackage{booktabs}
\usepackage{multirow}
\usepackage{multicol}
\usepackage{rotating}
\usepackage{subfig}
\usepackage{adjustbox}
\usepackage[utf8]{inputenc}
\definecolor{hidden-draw}{RGB}{205, 44, 36}
\definecolor{hidden-blue}{RGB}{194,232,247}
\definecolor{hidden-orange}{RGB}{243,202,120}
\definecolor{hidden-yellow}{RGB}{255,229,204}
\definecolor{hidden-red}{RGB}{255,204,204}
\usepackage{microtype}
\usepackage{xcolor,colortbl}
\usepackage{marvosym}
\usepackage{bm,amsmath}
\usepackage{multirow}
\usepackage{tabularx}
\usepackage{rotating}
\usepackage{makecell}
\usepackage{xspace,mfirstuc,tabulary}
\usepackage{hyperref}
\usepackage{color}
\usepackage{tikz}
\usepackage{graphicx}
\usepackage{pifont}
\usepackage[edges]{forest}
\definecolor{hidden-draw}{RGB}{20,68,106}
\definecolor{hidden-pink}{RGB}{255,245,247}

\theoremstyle{thmstyleone}%
\theoremstyle{thmstyletwo}%

\theoremstyle{thmstylethree}%

\begin{document}

\title[Article Title]{On-Device Recommender Systems: A Comprehensive Survey}


\author*[1]{\fnm{Hongzhi} \sur{Yin}}\email{h.yin1@uq.edu.au}
\equalcont{These authors contributed equally to this work.}

\author[2]{\fnm{Liang} \sur{Qu}}\email{l.qu@ecu.edu.au}
\equalcont{These authors contributed equally to this work.}

\author[1]{\fnm{Tong} \sur{Chen}}\email{tong.chen@uq.edu.au}

\author[1]{\fnm{Wei} \sur{Yuan}}\email{w.yuan@uq.edu.au}

\author[1]{\fnm{Ruiqi} \sur{Zheng}}\email{ruiqi.zheng@uq.net.au}

\author[1]{\fnm{Jing} \sur{Long}}\email{jing.long@uq.edu.au}

\author[1]{\fnm{Xin} \sur{Xia}}\email{x.xia@uq.edu.au}

\author[3]{\fnm{Yuhui} \sur{Shi}}\email{shiyh@sustech.edu.cn}

\author[4]{\fnm{Chengqi} \sur{Zhang}}\email{chengqi.zhang@polyu.edu.hk}

\affil*[1]{\orgname{The University of Queensland}, \orgaddress{\city{Brisbane}, \country{Australia}}}

\affil[2]{\orgname{Edith Cowan University}, \orgaddress{\city{Perth}, \country{Australia}}}

\affil[3]{\orgname{Southern University of Science and Technology}, \orgaddress{\city{Shenzhen}, \country{China}}}

\affil[4]{\orgname{The Hong Kong Polytechnic University}, \orgaddress{\city{Hongkong}, \country{China}}}


\abstract{Recommender systems have been widely deployed in various real-world applications to help users identify content of interest from massive amounts of information. 
Traditional recommender systems work by collecting user-item interaction data in a cloud-based data center and training a centralized model to perform the recommendation service. However, such cloud-based recommender systems (CloudRSs) inevitably suffer from excessive resource consumption, response latency, as well as privacy and security risks concerning both data and models. 
Recently, 
driven by the advances in storage, communication, and computation capabilities of edge devices, 
there has been a shift of focus from CloudRSs to on-device recommender systems (DeviceRSs), which leverage the capabilities of edge devices to minimize centralized data storage requirements, reduce the response latency caused by communication overheads, and enhance user privacy and security by localizing data processing and model training. Despite the rapid rise of DeviceRSs, there is a clear absence of timely literature reviews that systematically introduce, categorize and contrast these methods. 
To bridge this gap, we aim to provide a comprehensive survey of DeviceRSs, covering three main aspects:
(1) the deployment and inference of DeviceRSs, exploring how large recommendation models can be compressed and utilized within resource-constrained on-device environments; (2) the training and update of DeviceRSs, discussing how local data can be leveraged for model optimization on the device side; (3) the security and privacy of DeviceRSs, unveiling their potential vulnerability to malicious attacks and defensive strategies to safeguard these systems. 
Furthermore, we provide a fine-grained and systematic taxonomy of the methods involved in each aspect, followed by a discussion regarding challenges and future research directions. 
This is the first comprehensive survey on DeviceRSs that covers a spectrum of tasks to fit various needs. We believe this survey will help readers understand the current research status in this field, equip them with relevant technical foundations, and stimulate new research ideas for developing DeviceRSs.}




\maketitle
\section{Statement and Declarations}

\subsection{Availability of data and materials}
This is a survey article and does not involve any original experimental data or materials.

\subsection{Competing interests}
The authors declare that they have no competing interests.

\subsection{Funding}
The Australian Research Council partially supports this work under the streams of Future Fellowship (Grant No. FT210100624), Discovery Early Career Researcher Award (Grants No. DE230101033), the Discovery Project (Grant No. DP240101108 and DP240101814), and the Linkage Projects (Grant No. LP230200892 and LP240200546).

\subsection{Authors' contributions}
Hongzhi Yin proposed the original idea for the survey, led the design of the taxonomy, and was responsible for the overall revision of the manuscript.\\
Liang Qu contributed to writing the \textit{Training and Updating} sections.\\
Tong Chen led the classification of \textit{Deployment and Inference} approaches.\\
Ruiqi Zheng contributed to writing the \textit{Deployment and Inference} section.\\
Jing Long contributed to writing the \textit{Training and Updating} section.\\
Xin Xia contributed to writing the \textit{Deployment and Inference} section.\\
Yuhui Shi contributed to the revision and quality improvement of the manuscript.\\
Chengqi Zhang contributed to the revision and quality improvement of the manuscript.\\

\subsection{Acknowledgements}
Not applicable.

\section{Introduction}\label{sec1}

\begin{figure*}
    \centering
    \includegraphics[width=1\linewidth]{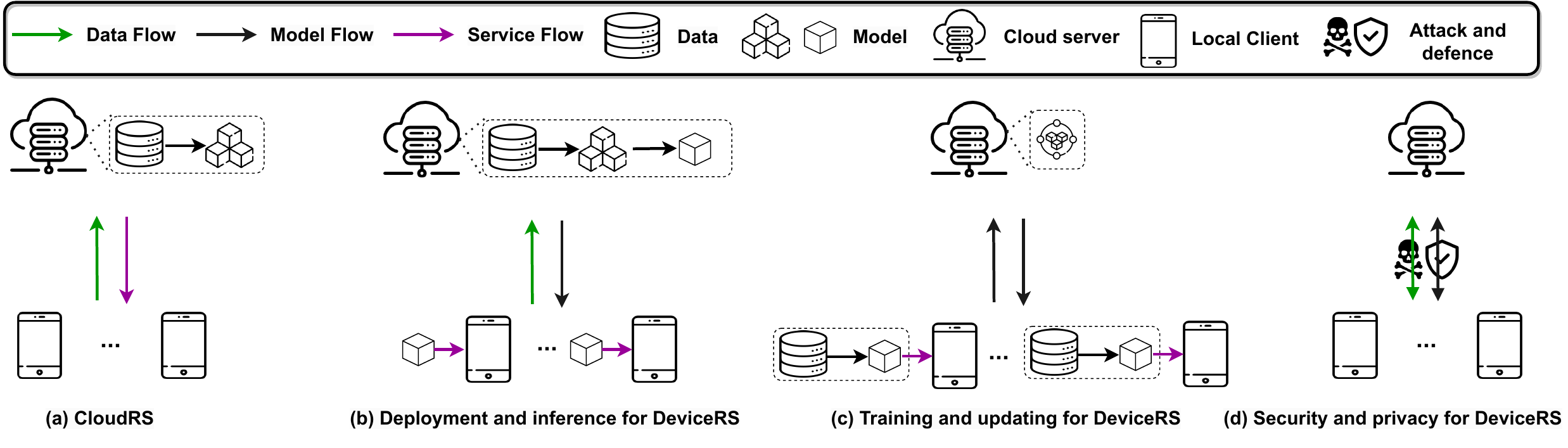}
    \caption{The illustration of CloudRS and three types of DeviceRSs.}
    \label{fig:CloudRSvsDeviceRS}
    
\end{figure*}

Recommender systems (RSs), as an important technology to help online users efficiently pinpoint relevant information from the Web, have seen a strong uptake in many application sectors such as e-commerce \cite{10.1145/3219819.3219823}, multimedia platforms \cite{covington2016deep}, location-based services~\cite{10.1145/2487575.2487608,10.1145/2873055,8013107,10.1145/2733373.2806339,wang2020next}, etc. Typically, most existing RSs \cite{guo2017deepfm,10.1145/3219819.3219823,ying2018graph} are deployed on the cloud server, where the recommendation model is trained and hosted in a centralized manner with data collected from users. The typical working mechanism of such cloud-based recommenders, termed \textit{CloudRSs}, is depicted in Figure \ref{fig:CloudRSvsDeviceRS} (a). However, given the intrinsic characteristics of this workflow, CloudRSs are inevitably subject to the following deficiencies: 
(1) \textbf{High Resource Consumption:} On the one hand, there is the substantial storage requirement for extensive data associated with users and items, including user-item interaction data, the features of users and items, and the parameters of the models, such as the weights of neural networks and user/item embeddings. On the other hand, the continuous updating and training of models to capture evolving user preferences also contribute substantially to resource consumption.
(2) \textbf{Unwarranted Response Latency:} Another inherent limitation of the CloudRS architecture is the response latency between the cloud and user devices due to the communication overhead. For example, devices need to receive results from the CloudRS for display, and this response time can be impaired by limited Internet connection or high-traffic periods. Additionally, many models rely on historical instead of real-time user data for training and predictions, which often results in failing to capture real-time changes in user preferences.
(3) \textbf{Security and Privacy:} Holding all user data in the data center could pose a risk to data security and privacy. In addition, with the release of data protection regulations such as GDPR\footnote{https://gdpr-info.eu/} from Europe, CCPA\footnote{https://oag.ca.gov/privacy/ccpa} from the USA, and PIPL\footnote{https://personalinformationprotectionlaw.com} from China, there is a strong demand from users for enhancing data privacy protection. Furthermore, CloudRSs are vulnerable to attacks that can manipulate their outputs, leading to the generation of false or biased results.

Recent years have witnessed the rapid technical development in edge devices that has uplifted their storage, communication, and computational capacity. As such, a brand new recommendation paradigm, namely on-device recommender systems (DeviceRSs), are proposed to transit all or a bulk part of the computation and storage responsibilities from the cloud server to users' end-devices, such as cellphones, tablets, and smart home speakers. Given DeviceRSs' capability of addressing the aforementioned issues by putting the model deployment and even training of recommendation models on the device side, there is also an emerging line of real-world applications such as Taobao's EdgeRec \cite{gong2020edgerec}, Google's TFL Recommendation\footnote{https://www.tensorflow.org/lite/examples/recommendation/overview}, and Kuaishou's short video recommendation on mobile devices \cite{gong2022real}. Specifically, as systematically outlined in Figure \ref{fig:taxonomy}, existing DeviceRSs could be grouped into three broad categories, which are summarized in the following three paragraphs.

\tikzstyle{my-box}=[
    rectangle,
    draw=hidden-draw,
    rounded corners,
    text opacity=1,
    minimum height=2em,
    minimum width=5em,
    inner sep=2pt,
    align=center,
    fill opacity=.5,
    line width=0.8pt,
]
\tikzstyle{leaf}=[my-box, minimum height=1.5em,
    text=black, align=left,font=\huge,
    inner xsep=4pt,
    inner ysep=6pt,
    line width=1pt,
]

\begin{figure*}[th!]
    \centering
    \resizebox{1\textwidth}{!}{
        \begin{forest}
            forked edges,
            for tree={
                grow=east,
                reversed=true,
                anchor=base west,
                parent anchor=east,
                child anchor=west,
                base=left,
                font=\Large,
                rectangle,
                draw=hidden-draw,
                rounded corners,
                align=left,
                minimum width=4em,
                edge+={darkgray, line width=1pt},
                s sep=7pt,
                inner xsep=2pt,
                inner ysep=3pt,
                line width=0.8pt,
                ver/.style={rotate=90, child anchor=north, parent anchor=south, anchor=center},
            },
            where level=1{text width=6em,font=\Large,}{},
            where level=2{text width=6em,font=\Large,}{},
            where level=3{text width=9em,font=\normalsize,}{},
            where level=4{text width=5em,font=\tiny,}{},
            [
                DeviceRSs,fill=hidden-yellow!70
                [
                    Deployment and Inference \\ (Section \ref{section:deployandinference}),text width=18em,fill=hidden-blue!70
                    [
                     Binary Code-based Methods\\ (Section \ref{section:DI_BC}),text width=21em,fill=hidden-blue!70
                           [
                                CH~\cite{liu2014collaborative}\text{,}
                                PPH~ \cite{zhang2014preference}\text{,}
                                DCF~\cite{zhang2016discrete}\text{,}\\
                                DCMF~\cite{lian2017discrete}\text{,}
                                CIGAR~ \cite{kang2019candidate}\text{,} 
                                HashGNN~\cite{tan2020learning}\text{,}\\
                                PTQ~\cite{guan2019post}\text{,}
                                StocQ~ \cite{xu2021agile}\text{,} 
                                ALPT~\cite{li2023adaptive},
                                leaf, font=\Large, text width=33em,fill=hidden-blue!70
                            ]
                    ]
                    [
                     Embedding Sparsification \\ Methods (Section \ref{section:DI_ES}),text width=21em,fill=hidden-blue!70
                           [
AMTL~\cite{yan2021learning}\text{,}PEP~\cite{liu2021learnable}\text{,}CIESS~\cite{qu2023continuous}\\
,
                                leaf, font=\Large, text width=33em,fill=hidden-blue!70
                            ]
                    ]
                    [
                    Variable Size Embedding  \\ Methods  (Section \ref{section:DI_VSE}),text width=21em,fill=hidden-blue!70
                           [
MDE~\cite{ginart2021mixed}\text{,}
DNIS~\cite{cheng2020differentiable}\text{,}
AutoEmb~\cite{zhao2020autoemb}\text{,}\\
AutoDim~\cite{zhao2021autodim}\text{,}
ESAPN~\cite{liu2020automated}\text{,}\\
RULE~\cite{chen2021learning}\text{,}
DeepLight~\cite{deng2021deeplight}\text{,}
SSEDS~\cite{qu2022single}
,
                                leaf, font=\Large, text width=33em,fill=hidden-blue!70
                            ]
                    ]
                    [
                     Compositional Embedding \\ Methods (Section \ref{section:DI_CE}),text width=21em,fill=hidden-blue!70
                           [
QRT~\cite{shi2020compositional}\text{,}
MEmCom~\cite{MLSYS2022_72988287}\text{,}
BH~\cite{10.1145/3459637.3482065}\text{,}\\
DHE~\cite{kang2021learning}\text{,}
LLRec~\cite{wang2020next}\text{,}
TT-Rec~\cite{yin2021tt}\text{,}\\
ODRec~\cite{xia2022device}\text{,}
LightRec~\cite{lian2020lightrec}\text{,}
EODRec~\cite{xia2023efficient}
,
                                leaf, font=\Large, text width=33em,fill=hidden-blue!70
                            ]
                    ]
                    [
                    Sustainable Deployment \\ Methods (Section \ref{section:DI_SD}),text width=21em,fill=hidden-blue!70
                           [
                                MoMoDistill~\cite{yao2021device}\text{,}
                                ODUpdate~\cite{xia2023towards}
                                ,leaf, font=\Large, text width=33em,fill=hidden-blue!70
                            ]
                    ]                     
                ]
                [
                    Training and updating  \\ (Section \ref{Section:ondevicelearning}),text width=18em,fill=hidden-red!70
                    [
                     Federated Recommendation \\ Methods (Section \ref{section:FedRS}),text width=21em,fill=hidden-red!70
                           [
                                FCF~\cite{ammad2019federated}\text{,} 
                                FedMF~ \cite{chai2020secure}\text{,} SeSoRec~\cite{chen2020secure}\text{,} \\
                                FedNewsRec~\cite{qi2020privacy}\text{,} MetaMF~\cite{10.1145/3397271.3401081}\text{,}
                                FedeRank~\cite{anelli2021federank}\\   S$^{3}$Rec~\cite{cui2021exploiting}\text{,}
                                FMF-LDP~\cite{minto2021stronger}\text{,}
                                FedRec~\cite{lin2020fedrec}\text{,}\\
                                FedRec++~\cite{liang2021fedrec++}\text{,}
                                FCF-BTS \cite{10.1145/3460231.3474257}\text{,}
                                FedCT \cite{10.1145/3404835.3462825}\text{,}\\
                                PrivRec \cite{10.1007/s00778-021-00700-6}\text{,}
                                HPFL \cite{10.1145/3442381.3449926}\text{,}
                                FCMF~\cite{yang2021fcmf}\text{,}\\
                                PriCDR \cite{chen2022differential}\text{,}
                                F2MF \cite{liu2022fairness}\text{,}
                                FedNCF \cite{10.1016/j.knosys.2022.108441}\text{,}\\
                                FedPerGNN \cite{wu2022federated}\text{,}
                                FeSoG \cite{10.1145/3501815}\text{,}
                                PerFedRec \cite{10.1145/3511808.3557668}\text{,}\\
                                FedCTR \cite{10.1145/3506715}\text{,}
                                FedCDR~\cite{10.1145/3511808.3557320}\text{,}
                                FMSS \cite{10.1145/3548456}\text{,}
                                \\FedDSR \cite{9626622}\text{,}
                                PrivateRec \cite{liu2023privaterec}\text{,}
                                FPPDM \cite{liufederatedP}\text{,}\\
                                VFUCB \cite{cao2023privacy}\text{,}
                                SemiDFEGL~\cite{10.1145/3543507.3583337}\text{,} FedFast~\cite{10.1145/3394486.3403176}\text{,}\\
                                ReFRS \cite{10.1145/3560486}\text{,}
                                PFedRS \cite{zhang2023dual}\text{,}
                                LightFR \cite{zhang2023lightfr}\text{,}\\
                                CPF-POI \cite{10241965}\text{,}
                                FedCORE \cite{10443503}\text{,}
                                GPFedRec \cite{10.1145/3637528.3671702}\text{,}\\
                                FedRAP \cite{li2024federated}\text{,}
                                CoFedRec \cite{10.1145/3589334.3645626}\text{,}
                                FedHGNN \cite{10.1145/3589334.3645693}\text{,}\\
                                PoisonFRS \cite{10.1145/3589334.3645492}\text{,}
                                DGFedRS \cite{10.1145/3688570}\text{,}
                                PTF-FSR \cite{10.1145/3708344}\text{,}\\
                                FedDAE \cite{li2025personalized}
                                ,
                                leaf, font=\Large, text width=33em,fill=hidden-red!70
                            ]
                    ]
                    [
                     Decentralized Recommendation \\ Methods (Section \ref{section:DecRecs}),text width=21em,fill=hidden-red!70
                           [
                                DMC~\cite{wang2015DRS}\text{,}
                                PRW~\cite{kermarrec2010RW}\text{,}
                                DMF~\cite{chen2018DMF}\text{,}\\
                                Dec-GS~\cite{ling2012DLMC}\text{,}           
                                DRMF~\cite{hegedHus2019decentralized}\text{,}
                                DANOS~\cite{defiebre2020decentralized}\text{,}\\
                                DANOS*~\cite{defiebre2022human}\text{,}
                                CUPDMRS~\cite{beierle2019collaborating}\text{,}
                                DPMF~\cite{yang2022dpmf}\text{,}\\
                                NRDL~\cite{an2024nrdl}\text{,}
                                DGREC~\cite{zheng2023decentralized}\text{,}
                                DCLR~\cite{long2023decentralized}\\
                                MAC~\cite{long2023model}
                                ,leaf, font=\Large, text width=33em,fill=hidden-red!70
                            ]
                    ]
                    [
                     On-device Finetuning Methods \\ (Section \ref{section:DeviceRSfinetuning}),text width=21em,fill=hidden-red!70
                           [
                                PrivRec~\cite{10.1007/s00778-021-00700-6}\text{,}
                                MPDA~\cite{yan2022device}\text{,}
                                ODPF~\cite{mairittha2020improving}\text{,}\\
                                DCCL-GS~\cite{yao2021device}
                                ,leaf, font=\Large, text width=33em,fill=hidden-red!70
                            ]
                    ]                    
                ]
                [
                    Security and Privacy \\ (Section \ref{section:securityandprivacy}),text width=18em,fill=hidden-pink!70
                    [
                     Privacy Risks and\\ Countermeasures (Section \ref{section:privacyrisksandcountermeasures}),text width=21em,fill=hidden-pink!70
                           [
                                FedGNN~\cite{wu2021fedgnn}\text{,} FedMF~\cite{chai2020secure}\text{,} IMIA~\cite{yuan2023interaction}\text{,} \\
                                Fed-APM~\cite{zhang2023comprehensive}\text{,} FRU~\cite{yuan2023federated}\text{,} FedRec~\cite{lin2020fedrec}\text{,} \\FedRec++~\cite{liang2021fedrec++}\text{,} FeSoG~\cite{10.1145/3501815}\text{,} FR-FMSS~\cite{10.1145/3460231.3478855}\text{,}\\ FMSS~\cite{10.1145/3548456}\text{,} FedNewsRec~\cite{qi2020privacy}\text{,} PerFedRec~\cite{10.1145/3511808.3557668}\text{,} \\perFedRec++~\cite{luo2023perfedrec++}\text{,} FedPoiRec~\cite{perifanis2023fedpoirec}\text{,} \\SharedMF~\cite{ying2020shared}\text{,} PTF-FedRec~\cite{yuan2023hide},
                                leaf,font=\Large, text width=33em,fill=hidden-pink!70
                            ]
                    ]
                    [
                     Poisoning Attacks and\\ Countermeasures  (Section \ref{section:robustnessrisksandcountermeasures}),text width=21em,fill=hidden-pink!70
                           [
                                FSAD~\cite{jiang2020detection}\text{,} FedAttack~\cite{wu2022fedattack}\text{,} PipAttack~\cite{zhang2022pipattack}\text{,} \\FedRecAttack~\cite{rong2022fedrecattack}\text{,} A-hum and A-ra~\cite{rong2022poisoning}\text{,} \\PSMU~\cite{yuan2023manipulating}\text{,} PSMU(V)~\cite{yuan2023manipulating1}\text{,} \\ClusterAttack~\cite{yu2023untargeted}\text{,} A-FRS~\cite{chen2020robust}
                                ,leaf, font=\Large, text width=33em,fill=hidden-pink!70
                            ]
                    ]                 
                ]                  
            ]
        \end{forest}}
        \vspace{+10pt}
    \caption{Taxonomy of DeviceRSs.}
    \label{fig:taxonomy}
\end{figure*}

\textbf{On-Device Deployment and Inference for DeviceRSs} aim to deploy a lightweight recommendation model on resource-constrained devices, as shown in Figure \ref{fig:CloudRSvsDeviceRS} (b). This allows for rapid on-device model inference, alleviating resource consumption and response latency issues. The primary technical challenge associated with such type of methods is how to retain the performance of the original model as much as possible while compressing it. The existing methods can be further categorized into several distinct types including Binary Code-based Methods \cite{zhang2016discrete,zhang2017discrete}, Embedding Sparsification Methods \cite{liu2021learnable,qu2023continuous}, Compositional Embedding Methods \cite{wang2020next,shi2020compositional,lian2020lightrec,xia2022device,xia2023efficient}, Variable Size Embedding Methods \cite{liu2020automated,chen2021learning,kang2021learning}, and Sustainable Deployment methods \cite{xia2023towards}.

\textbf{Training and Update for DeviceRSs} involve shifting the training process to the device side by leveraging locally stored data, as shown in Figure \ref{fig:CloudRSvsDeviceRS} (c), thereby alleviating security and privacy concerns associated with the data uploading process. Additionally, local model updates can promptly capture changes in user preferences. However, due to the inherent data sparsity characteristics of RSs, the training data on each device is usually limited, making it challenging to achieve good performance through local data training alone. In order to address this issue, some methods in this field involve a central server that coordinates a group of devices for collaborative training between server and devices (e.g., federated RSs \cite{sun2022survey,yang2020federated,alamgir2022federated}), and others adopt the P2P protocol to perform decentralized collaborative training among devices \cite{long2023decentralized,10.1145/3543507.3583337,long2023model}. Recently, some research has focused on adapting the pre-trained large general models on the cloud server to the devices by fine-tuning using local data \cite{10.1007/s00778-021-00700-6,yao2021device,yan2022device}.

\textbf{Security and Privacy for DeviceRSs} aim to protect both users and on-device models from potential malicious attacks, as shown in Figure \ref{fig:CloudRSvsDeviceRS} (d). 
On the one hand, when local models contribute to a shared learning process, there is a risk of leaking sensitive user data. Even though direct user data might not be shared, sophisticated analysis of model updates can potentially reveal personal information. This is especially true in scenarios where the model updates reflect unique user behaviors or preferences in the context of recommendation \cite{yuan2023interaction,chai2022efficient,yuan2023federated,zhang2023comprehensive}.
On the other hand, since the system relies on local data from user devices for training and updating models, there is a possibility that an attacker could manipulate this data or the model's learning process. For instance, an attacker might introduce poisoned data or adversarial inputs into their local dataset, which could lead to the generation of inaccurate or biased recommendations \cite{zhang2022pipattack,wu2022fedattack,yuan2023manipulating}. 

There are a number of literature reviews on recommender systems and on-device learning. For example, Gao \textit{et al.} \cite{10.1145/3488560.3501396} and Wu \textit{et al.} \cite{10.1145/3535101} provide comprehensive reviews that focus on graph neural networks (GNNs) based recommender systems. Zheng \textit{et al.} \cite{zheng2023automl} and Tang \textit{et al.} \cite{tang2023automl} introduce a range of work on recommender systems from an automated machine learning perspective. Yu \textit{et al.} \cite{yu2022self} build a systematic taxonomy on self-supervised recommendations. Li \textit{et al.} \cite{li2023embedding} review methods for embedding compression in recommender systems. Zhang \textit{et al.} \cite{zhang2024survey} offer a comprehensive survey on Point-of-Interest (POI) recommender systems. Himeur \textit{et al.} \cite{himeur2022latest} and Rafi \textit{et al.} \cite{rafi2024fairness} discuss the privacy preserving techniques in recommender systems. Wang \textit{et al.} \cite{wang2024trustworthy} present a overview of trustworthy recommender systems. 
On the other hand, Dhar \textit{et al.} \cite{10.1145/3450494} summarizes a set of works on on-device learning from the perspective of algorithms and learning theories, and Zhou \textit{et al.} \cite{9366901} presents the on-device learning techniques from a software and hardware synergy perspective. However, these existing surveys treat recommender systems and the on-device learning in separate ways, so there is a timely demand for a comprehensive review that focuses on techniques in on-device recommender systems. The most similar works to ours are surveys on federated recommender systems \cite{sun2022survey,yang2020federated,alamgir2022federated,asad2023comprehensive,harasic2024recent,yang2025survey,10.1145/3708982}, but they only include limited works from the on-device training (i.e., federated learning) perspective, thereby missing the recent development of on-device deployment and attacking for recommendation.

Due to the sheer volume of work in the field of DeviceRSs, it is impractical to cover every paper in our survey. Our criteria for selecting papers are based on those published in top-tier journals and conferences recognized in this field, including but not limited to TKDE (IEEE Transactions on Knowledge and Data Engineering), TOIS (ACM Transactions on Information Systems), and notable conferences like KDD (ACM SIGKDD Conference on Knowledge Discovery and Data Mining), SIGIR (ACM SIGIR Conference on Research and Development in Information Retrieval), WWW (The Web Conference), WSDM (ACM Conference on Web Search and Data Mining) and RecSys (ACM Conference on Recommender Systems).
In addition to these peer-reviewed works, we also consider non-peer-reviewed valuable works, particularly those available on platforms like arXiv. However, our focus here is on those works that have gained significant recognition in the academic community and are deemed to be pioneering efforts in this field.  

Overall, this paper aims to present a up-to-date and comprehensive review that covers a broader range of methods for DeviceRSs. The contributions of this survey are summarized as follows:
\begin{itemize}
    \item \textbf{Comprehensive Review:} To our best knowledge, this is the first comprehensive survey of on-device recommender systems (DeviceRSs), offering an extensive overview of DeviceRSs technologies and their applications across various recommendation tasks. It includes an exploration of diverse approaches within DeviceRSs, highlighting their adaptability and functionality in different scenarios.
    \item \textbf{Systematic Taxonomy :} We propose a novel systematic taxonomy of the existing DeviceRS methods, focusing on three critical aspects: the deployment and inference for DeviceRSs, the training and updating for DeviceRSs, and the privacy and security for DeviceRSs. This taxonomy provides a structured insight into the state-of-the-art methods and their operational frameworks.
    \item \textbf{Challenges and Future Directions:} We dive into a detailed discussion regarding the potential challenges that DeviceRS methods currently face and propose future directions for research in this domain. This includes identifying gaps in the existing literature, suggesting improvements for current methodologies, and forecasting emerging trends that could influence the evolution of DeviceRSs.
\end{itemize}

The remaining sections of this survey are organized as follows: Section \ref{sec:CloudRS} introduce preliminary concepts for traditional CloudRSs and motivations of DeviceRSs. Sections \ref{section:deployandinference}, \ref{Section:ondevicelearning}, and \ref{section:securityandprivacy} dive into various aspects of DeviceRSs, covering deployment and inference, on-device training and updating, and privacy and security, respectively. Section \ref{sec:metrics} presents a set of commonly used evaluation metrics. Section \ref{section:challengesanddirections} highlights potential challenges and future research directions, followed by a conclusion in Section \ref{section:conclusion}.

\section{Preliminary}\label{sec:CloudRS}
Before we dive into the various types of DeviceRSs, we first provide relevant definitions of recommender systems and then discuss the motivations for shifting CloudRSs to DeviceRSs.

Recommender systems (RSs) have emerged as a popular research topic in both academia and industry over the past decade, aiming at assisting users in selecting items that align with their preferences from a vast array of options like news, videos, and products, thereby alleviating the issue of information overload. Generally, RSs perform recommendation services by collecting user data $\mathcal{D}=\{\mathcal{U},\mathcal{I},\mathcal{R},\mathcal{X}\}$ to train a recommendation model $RS(\textbf{P},\textbf{Q},\Theta)$ for a given recommendation task $T$, which is defined as
$RS(\textbf{P},\textbf{Q},\Theta): \mathcal{D}=\{\mathcal{U},\mathcal{I},\mathcal{R},\mathcal{X}\} \rightarrow T$,
where $\mathcal{U}$,$\mathcal{I}$,$\mathcal{R}$, and $\mathcal{X}$ denote the user set, item set, user-item interactions, and features associated with users/items, respectively. $\textbf{P}$, $\textbf{Q}$, and $\Theta$ represent user representations, item representations, and model parameters, respectively.
CloudRSs can be further classified into different categories based on the data, models, and recommendation tasks:
\begin{itemize}
    \item Based on different data types, RSs can be categorized into collaborative filtering-based methods, which mainly utilize user-item interaction data $\mathcal{D}=\{\mathcal{U},\mathcal{I},\mathcal{R}\}$, and content-based methods (i.e., $\mathcal{D}=\{\mathcal{U},\mathcal{I},\mathcal{R},\mathcal{X}\}$) that integrate side information associated users and items (e.g., user/item features) with user-item interaction data to improve recommendation performance.
    \item Depending on the recommendation model employed, RSs can be further divided. Matrix factorization-based methods (i.e., $RS(\textbf{P},\textbf{Q})$) \cite{koren2009matrix} decompose the user-item interaction matrix into two low-dimensional matrices for users and items. Each row (or column) vector in these matrices represents the user or item embeddings. With the rise of deep learning, many methods employ DNNs parameterized by parameters $\Theta$ to learn user/item embeddings to capture the nonlinear relationships between users and items, i.e., $RS(\textbf{P},\textbf{Q},\Theta)$ \cite{covington2016deep}. Additionally, since user-item interaction data can naturally be modeled as a user-item bipartite graph, some methods apply various graph neural network techniques to this bipartite graph to learn user/item embeddings \cite{10.1145/3535101}.
    \item Based on different recommendation tasks, RSs can also be modeled for optimizing various objectives. For regression tasks, like rating prediction \cite{lei2016rating}, the system predicts specific ratings that a user might give to items. For ranking tasks, such as top-$k$ recommendation \cite{hidasi2018recurrent}, it predicts the top $k$ items most relevant to a given user. Lastly, for classification tasks like CTR (Click-Through Rate) prediction \cite{xu2021agile,guo2017deepfm}, RSs predict whether a user will click on certain items. 
\end{itemize}

\section{Deployment and Inference for DeviceRS}\label{section:deployandinference}

\begin{figure*}
\setlength{\abovecaptionskip}{1pt}
    \centering
    \includegraphics[width=1\linewidth]{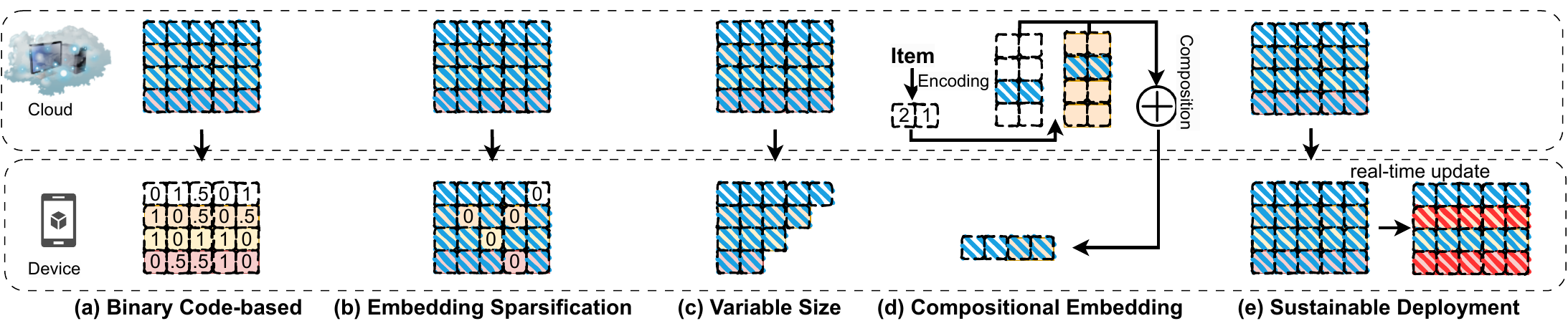}
    \caption{The illustration of five types of methods within deployment and inference for DeviceRSs.}
    \label{fig:DeviceRSCompression}
\end{figure*}

In this section, we will introduce the deployment and inference methods for DeviceRSs, aimed at compressing large recommender system models to accommodate the memory limits of device-side platforms. The current trend involves transitioning models from well-resourced cloud environments to the more memory-restricted device side. In contrast to Natural Language Processing and Computer Vision, which utilize deep neural networks with substantial layers, recommender systems typically opt for shallower models. The majority of parameters in these systems are concentrated in the embedding tables. 
While some model compression methods \cite{sun2020generic,yuan2020parameter,wang2021stackrec,liu2022one,chen2021user} might lack explicit mention as on-device models in their original papers, they inherently possess the potential for deployment on devices. Therefore, we will also introduce a range of general recommendation model compression methods \cite{yan2021learning,zhang2014preference,joglekar2020neural}, highlighting their capabilities and potential benefits for deployment on devices.
As depicted in Figure \ref{fig:DeviceRSCompression}, deployment and inference methods for DeviceRSs primarily focus on embedding table compression. This typically includes: (1) Binary code-based methods involve converting embedding weights, typically stored in the format of FP32 (floating point 32-bit), to low-precision formats with fewer bits; (2) Embedding sparsification methods entail masking less important parts of embedding tables with zeros; (3) Variable size embedding methods aim to search the appropriate embedding length for different features; (4) Compositional embedding methods share weights among the embedding tables; (5) Sustainable deployment methods focus on achieving real-time on-device model updates post-deployment.

\subsection{Binary Code-based Embedding}\label{section:DI_BC}

Specifically, concerning the bit width of weights, Binary Code-based DeviceRSs can be categorized as binarization and quantization.

\subsubsection{Binarization}

It compresses an embedding weight from full-precision as binary code of just one bit, which presents two advantages: they demand less memory for storage, and the computation of similarity (inner product) becomes notably efficient through the employment of Hamming distance. This technique aligns with the broader trend of deploying recommender models on memory-restricted device-side platforms.

CH \cite{liu2014collaborative} and PPH \cite{zhang2014preference} initially acquire compressed embeddings through binarization in two steps processes for Collaborative Filtering (CF) \cite{wu2022survey}. The procedure involves first learning embedding table with full precision solely without considering memory constraints. Subsequently, binarization is applied to these original embedding tables to obtain binary compressed embeddings.
Nevertheless, such a compression step does not incorporated with the learning procedure and is unable to be learned jointly with the embedding table as a training objective. This limitation may result in substantial and irreparable errors. To address this issue and mitigate potential accuracy loss, subsequent efforts have shifted towards end-to-end frameworks. These methods aim to optimize compressed binary embeddings within the learning phase, enhancing the model's ability to maintain accuracy while incorporating binary constraints.

For direct optimization, binary embeddings are seamlessly integrated into the model parameters and directly optimized through the training loss. DCF \cite{zhang2016discrete} employs a binary embedding table $\textbf{B} \in \{1,-1\}^{n \times d}$ alongside a original uncompressed embedding table $\textbf{E} \in \mathbb{R}^{n \times d}$. The loss of $(\textbf{B} - \textbf{E})$ is incorporated into the optimization. Specifically, updates for \textbf{B} are performed through Discrete Coordinate Descent, while updates for \textbf{E} are facilitated by Singular Value Decomposition. DCMF \cite{lian2017discrete} implements the same optimization strategy and extends the task from rating prediction to top-$k$ recommendation.

Differing from the direct optimization of binarization, indirect optimization involves generating the compressed embeddings from the original embeddings on the fly and optimizing them indirectly by optimizing the original full-precision embeddins. To address the challenge of zero gradients associated with the binary function (e.g., the sign function $sign(x)$), CIGAR \cite{kang2019candidate} implements a scaled hyperbolic tangent function $tanh(\alpha x)$, where $\lim_{\alpha \rightarrow \infty}tanh(\alpha x) = sign(x)$. 
An alternative approach is the Straight-Through Estimator (SET) \cite{courbariaux2015binaryconnect}. HashGNN \cite{tan2020learning} implements the STE variant of the $sign(x)$ function, optimizing \textbf{E} with the gradients of \textbf{B}. 


\subsubsection{Quantization}

To strike a balance between prediction accuracy and memory cost, quantization employs a multi-bit integer to represent each weight, mitigating accuracy drop associated with strict binarization. This process entails mapping an embedding weight consisting of 32-bit full-precision to a value within a set of quantized values.

In the context of post-training quantization (PTQ) \cite{guan2019post}, two quantization approaches are implemented. The first approach is uniform quantization, which involves upholding a range for each embedding to be quantized. The optimal range is determined through a greedy search. The second approach is non-uniform quantization, which involves grouping similar embeddings and employing clustering for each group to generate a codebook. The weights within each group are subsequently connected to the index value of the corresponding codebook. Recent works such as StocQ \cite{xu2021agile} and ALPT \cite{li2023adaptive} have introduced approaches to learning quantized weights from scratch. In contrast to directly quantizing full-precision weights in the forward manner and updating them with gradients estimated by STE, StocQ, and ALPT store weights in integer format during training to compress model memory size. Then de-quantized weights are taken as the input of the model, and after backward propagation, the model re-quantizes the weights back into integers. Given that the input of recommendation systems (e.g., one-hot vectors) is highly sparse, only a small portion of the embeddings will then be de-quantized by the model into floating-point values.

\subsection{Embedding Sparsification}\label{section:DI_ES}
Embedding sparsification methods, in contrast to binary code-based embedding methods that convert embedding weights to low-precision formats with fewer bits, identify less crucial embedding weights. Subsequently, these weights are masked with zeros using a sparsification technique.

\textbf{ATML} \cite{yan2021learning} suggests using an Adaptively-Masked Twins-based layer to reduce useless dimensions in the original embedding table through the mask vector. It incorporates feature frequency knowledge, employing l-AML and h-AML branches to handle low- and high-frequency instances, preventing dominance of parameters by high-frequency samples.

Learnable thresholds play an important role in learning the significance of parameters in the embedding matrix. Drawing inspiration from Soft Threshold Reparameterization, \textbf{PEP} \cite{liu2021learnable} dynamically prunes by adaptively reparameterizing the embedding matrix using $    \hat{\textbf{E}} = \mathcal{S}(\textbf{E},s) =  sign(\textbf{E})ReLU(abs(\textbf{E})-(1+e^{-s})^{-1})$, where $\hat{\textbf{E}}$ and $\textbf{E}$ denote the re-parameterized and original embedding matrices, respectively. The learnable threshold(s) $s$ are updated via gradient descent, and $sign(\cdot)$ represents the sign function. Furthermore, OptEmbed \cite{lyu2022optembed} not only employs learnable pruning thresholds for redundant embeddings but also implements a field-wise dimension mask for better performance.

Reinforcement learning offers another avenue for embedding sparsification. Unlike other RL-based methods with discrete choices, CIESS \cite{qu2023continuous} adopts an actor-critic paradigm, including the twin delayed deterministic policy gradient, to navigate a continuous search space, allowing flexible and arbitrary adjustment of embedding sizes.


\subsection{Variable Size Embedding}\label{section:DI_VSE}
Similar to embedding sparsification that assigns 0 to a certain embedding weight to shrink memory consumption, variable size embedding DeviceRSs methods
assign different sizes of embedding dimensions to each feature (field). It can not only save memory but also lead to better performance. Given the long-tail distribution observed in recommendation tasks \cite{park2008long}, assigning identical dimensions to both head and tail features is not rational. High-frequency features, typically following a long-tail distribution, require more weight to store the extensive information. Conversely, representing low-frequency features with excessive parameters could lead to overfitting because of limited training data.

One possible search strategy is based on human rules, where MDE \cite{ginart2021mixed} proposes a mixed-dimension embedding layer that dynamically learns the dimension of the embedding according to the popularity of features. 
Specifically, the MD layer is comprised of $F$ blocks, each matching to $F$ feature fields in the context of the CTR task. Every block instance is reshaped into two matrices: the projection matrix and the embedding matrix.
MDE initially establishes a block-level probability vector, and then the mixed embedding dimensions based on the popularity and the two matrices.

Motivated by the success of Neural Architecture Search (NAS) \cite{pham2018efficient}, a research line treats embedding optimization as a hyper-parameter optimization problem. This involves searching the needed embedding dimensions from a human-defined search space. NIS \cite{joglekar2020neural} learns vocabulary sizes and embedding dimensions for unique features. Building on this, Differential Neural Input Search (DNIS) \cite{cheng2020differentiable} enhances efficiency by introducing a differential framework, enabling continuous dimension selection instead of pre-defined discrete dimension sets. AutoEmb \cite{zhao2020autoemb}, an AutoML-based framework for sequential recommendations, dynamically selects embedding dimensions based on changing popularity for users and items in ranking prediction tasks. AutoDim \cite{zhao2021autodim} broadens this scope to recommendation tasks with multiple feature fields by assigning dimensions for different feature fields. ESAPN \cite{liu2020automated} dynamically searches embedding dimensions with reinforcement learning, employing a hard selection strategy, and diverging from the soft selection approach utilized by AutoEmb to effectively reduce storage space.

This problem can also be formulated as embedding pruning optimization, where pruning is applied to the entire embedding matrix using various strategies to automatically achieve variable-sized embeddings. Deeplight \cite{deng2021deeplight} addresses high-latency concerns in CTR prediction by proposing pruning for both embedding and DNN layers. This involves pruning weight matrices in the DNN component to eliminate connections, resulting in a sparse DNN with reduced computational complexity and accelerated training, Additionally, field pair interaction matrices are pruned for field pair selection. RULE \cite{chen2021learning} learns the full embedding table and searches for elastic embeddings for multiple groups of items using evolutionary search. As a once-for-all framework, it accommodates on-device recommenders with varying memory restrictions without training from scratch.
SSEDS \cite{qu2022single} presents embedding pruning in a single-shot manner. Initially, a recommendation model is pre-trained with the same embedding dimensions. Following this, a proposed criterion is employed, measuring the importance of each embedding dimension in a single forward-backward pass. This process results in salience scores assigned to each dimension to fit the memory budget.


\subsection{Compositional Embedding}\label{section:DI_CE}

Binary code-based methods decrease the bit count of embedding weight, while embedding sparsification and variable size embedding techniques shorten the embedding length. In contrast, compositional techniques share embedding weights among embedding tables, effectively reducing the overall parameter count. 

We extend the concept of compositional embedding and utilize a weight-sharing framework proposed by current methods \cite{li2023embedding}. Concretely, the generation of embedding is defined as $\textbf{e} = \sum / \bigcup^{s}_{i=1} \textbf{I} ^{i} \times \textbf{{T}}^{i}$. $ \sum / \bigcup$ represents summation or concatenation, $\textbf{I} ^{i}$ are meta-embeddings to determine shared vectors in $\textbf{{T}}$, and $\textbf{{T}}^{i}$ are meta-tables composed of shared vectors.

One kind of approach is hashing methods, which learn the meta-embeddings by employing hash functions to the original feature ID. Each feature is represented by an embedding $\textbf{e} = \textbf{I} \times \textbf{T}$, where $\textbf{T} \in \mathbb{R}^{m \times d}$,$m < n $,  $\textbf{I} = \mathbb{I}(ID\%m) \in \{0, 1\}^{m}$, and $\mathbb{I}$ is the one-hot encoding function. To reduce the collisions raised by mapping many features to the same embedding, multiple hash functions and index vectors with several hash function candidates are utilized for feature ID. For instance, QRT \cite{shi2020compositional} employs two meta-tables and utilizes reminder model and quotient model to learn dual index vectors, respectively. Following this, MEmCom \cite{MLSYS2022_72988287} implement two meta-tables ($\textbf{T}^{1} \in \mathbb{R}^{m \times d}$, $\textbf{T}^{2} \in \mathbb{R}^{n \times 1}$) and dual index vectors ($\textbf{I}^{1} = \mathbb{I}(ID\%m)$, $\textbf{I}^{2} = \mathbb{I}(ID)$), multiplying two meta-embeddings to obtain the final embedding to better distinguish multiple features. Furthermore, BH \cite{10.1145/3459637.3482065} clusters the full-precision feature ID into $s$ partitions and assigns them into $(ID_1,\cdot\cdot\cdot,ID_s)$. Every subgroup correlates to an index vector $\textbf{I}^{i} = \mathbb{I}(ID_i)$  and obtains a meta-embedding for sharing among all $\textbf{T}^{i}$.

Another research line is vector quantization, which learns the index vectors by the nearest-neighbor search to capture the similarity between features themselves. It's worth noting that in recent literature on vector quantization, the terms codeword and codebook are commonly used to refer to what we denote as meta-embedding and meta-table. For the original embedding $\textbf{e}$, vector quantization obtains the index vector by $\textbf{I} = \mathbb{I}(\arg \max_k sim(\textbf{e}, \textbf{T}_{k}))\in \{0,1\}^{m}$, where $\textbf{T}_k$ is the meta-embedding and $sim(\cdot)$ is the function to calculate similarity. For instance, LightRec \cite{lian2020lightrec} uses vector quantization to compress the item embedding tables and utilizes a pre-trained teacher model for efficiency. Furthermore, EODRec \cite{xia2023efficient} proposes a knowledge distillation paradigm with bidirectional self-supervised learning and saves meta-tables on local device for fast local inference. 

As for decomposition techniques, they implement soft selection through aggregating across all meta embeddings within a meta-table $\textbf{T}\in \mathbb{R}^{m \times d}$ with an index vector $\textbf{I} \in \mathbb{R}^{m}$ containing real values. Considering the extensive representational capacity of the index vectors, only one meta-table is satisfactory for addressing feature collisions. Every feature is represented by a unique index vector in  $\textbf{I}_M \in \mathbb{R}^{n \times m}$ when expressing the decomposition as $\textbf{E} = \textbf{I}_{M} \times \textbf{T}$. DHE \cite{kang2021learning} adopts a hash encoder $\mathcal{H}: \mathbb{N} \rightarrow \mathbb{R}^{m}$ for better  optimization and storage of $\textbf{I}_M$. To overcome the limitations of deterministic index vectors that cannot be optimized, DHE takes a step further by decomposing the meta-table $\textbf{T}$ into a DNN to enhance the expressive capacity of the meta-table.

Furthermore, tensor train decomposition (TTD) is employed for the decomposition of the embedding tables. The embedding table $\textbf{E} \in \mathbb{R}^{n \times d}$ is firstly reshaped into $\mathcal{E} \in \mathbb{R}^{(n_1 d_1)\times(n_2 d_2)\times\cdot\cdot\cdot\times(n_s d_s)}$, where $n= \prod^{s}_{i=1}n_i$ and $d= \prod^{s}_{i=1}d_i$, then, $\mathcal{E}$ is decomposed into $\mathcal{E} = \mathcal{G}_1 \times \mathcal{G}_2 \times \cdot\cdot\cdot\mathcal{G}_s$, where $\mathcal{G}_i \in \mathbb{R}^{r_{i-1} \times n_id_i\times r_i}$ is called TT-core and $r_i$
is called TT-rank. TT-Rec \cite{yin2021tt} stands out as the pioneer in applying Tensor-Train Decomposition (TTD) to recommendation tasks. This implementation involves optimized kernels specifically designed for TTD on embedding tables. On a similar note, LLRec \cite{wang2020next} utilizes TTD on embedding tables with a focus on preserving prediction accuracy through knowledge distillation. Facing the challenge of balancing the trade-off between under-expressive embedding approximations due to a small TT-rank in TTD and the sacrifice of model efficiency with a larger TT-rank, ODRec \cite{xia2022device} introduces semi-tensor product-based tensor train decomposition (STTD) as an enhancement to TTD for extremely compressing the embedding table. This adjustment enables TTD to perform tensor multiplication across varying ranks, effectively addressing concerns regarding efficiency and effectiveness.


\subsection{Sustainable Deployment Methods}\label{section:DI_SD}
The widespread adoption of compression techniques has offered solutions for deploying recommendation models on devices. However, a notable challenge has emerged concerning the sustainable implementation of on-device recommendations. The continuous evolution of user interests necessitates ongoing efforts in maintaining recommender systems. With millions of user interactions generated incessantly, this task becomes particularly demanding. However, device models can not achieve real-time updates for the lack of computing ability and storage space. One strategy is making cloud-device communication efficient where DeviceRSs can still benefit from the cloud. MoMoDistill \cite{yao2021device} first studies Device-Cloud Collaboration Learning framework to benefit both sides jointly. MoMoDistill reduces the parameter size of the cloud model to fit devices and then enhances the cloud modeling via the knowledge distillation sourced from the personalized device models. ODUpdate \cite{xia2023towards} proposes to compress model updates over time by encoding the model update into the compositional codes of items. Besides, they also designed two updating methods based on stack and queue, achieving different kinds of minimal communication costs.

\section{Training and update for DeviceRSs}\label{Section:ondevicelearning}
In this section, we will introduce the training and update methods for DeviceRSs, which means that a part or all of the model training and updating process is put on the device side,  thereby alleviating security and privacy concerns and enhancing the ability to capture changes in user preferences. Specifically, due to the
inherent data sparsity of user-item interactions, the training data on each device is usually limited, making it challenging to achieve good performance through local data training alone. Consequently, methods in this field typically involve a central server that coordinates a group of devices in different manners including (1) Federated recommendation methods that typically rely on a central server to coordinate model training via device-to-server communications; (2) Decentralized recommendation methods that typically perform device-to-device collaborative training among neighbors assigned from the server; (3) On-device finetuning methods that typically leverage local data to refine the pre-trained models from the server side. We elaborate these three types of methods separately in the following.

\subsection{Federated Recommender Systems}\label{section:FedRS}

\begin{table}[htbp]
  \tiny  
  \centering
  \caption{Summary of main FedRSs with their addressed tasks and types of participants.}
    \begin{tabular}{lrlll}
    \toprule
    \textbf{Method} & \multicolumn{1}{l}{\textbf{Year}} & \textbf{Venue} & \textbf{Task} & \textbf{Type} \\
    \midrule
    \midrule
    FCF \cite{ammad2019federated} & 2019  & ArXiv & Top-$k$ recommendation & Cross-client \\
    FedMF \cite{chai2020secure} & 2020  & IEEE Intelligent Systems & Rating prediction & Cross-client \\
    SeSoRec \cite{chen2020secure} & 2020  & ECAI 2020 & Rating prediction  & Cross-platform \\
    FedFast \cite{10.1145/3394486.3403176} & 2020  & KDD 2020 & Top-$k$ recommendation & Cross-client \\
    FedNewsRec \cite{qi2020privacy} & 2020  & EMNLP 2020 & Top-$k$  recommendation & Cross-client \\
    MetaMF \cite{10.1145/3397271.3401081} & 2020  & SIGIR 2020 & Rating prediction & Cross-client \\
    FedeRank \cite{anelli2021federank} & 2021  & ECIR 2021 & Top-$k$ recommendation & Cross-client \\
    S\^{3}Rec \cite{cui2021exploiting} & 2021  & NeurIPS 2021 & Rating prediction  & Cross-platform \\
    FMF-LDP \cite{minto2021stronger} & 2021  & RecSys 2021 & Top-$k$ recommendation & Cross-client \\
    FedRS \cite{lin2020fedrec} & 2021  & IEEE Intelligent Systems & Rating prediction & Cross-client \\
    FedRS++ \cite{liang2021fedrec++} & 2021  & AAAI 2021 & Rating prediction & Cross-client \\
    FCF-BTS \cite{10.1145/3460231.3474257} & 2021  & RecSys 2021 & Top-$k$ recommendation & Cross-client \\
    FedCT \cite{10.1145/3404835.3462825} & 2021  & SIGIR 2021 & Top-$k$ (cross-domain) & Cross-platform \\
    PrivRec \cite{10.1007/s00778-021-00700-6} & 2021  & VLDB Journal & Top-$k$ recommendation & Cross-platform \\
    HPFL \cite{10.1145/3442381.3449926} & 2021  & WWW 2021 & Rating prediction & Cross-platform \\
    FCMF \cite{yang2021fcmf} & 2021  & Knowledge-Based Systems & Rating prediction & Cross-platform \\
    PriCDR \cite{chen2022differential} & 2022  & WWW 2022 & Top-$k$ (cross-domain) & Cross-platform \\
    F2MF \cite{liu2022fairness} & 2022  & RecSys 2022 & Top-$k$ recommendation & Cross-client \\
    FedNCF \cite{10.1016/j.knosys.2022.108441} & 2022  & Knowledge-Based Systems & Top-$k$ recommendation & Cross-client \\
    FedPerGNN \cite{wu2022federated} & 2022  & Nature Communications & Rating prediction & Cross-client \\
    FeSoG \cite{10.1145/3501815} & 2022  & TIST  & Rating prediction & Cross-client \\
    PerFedRS \cite{10.1145/3511808.3557668} & 2022  & CIKM 2022 & Top-$k$ recommendation & Cross-client \\
    FedCTR \cite{10.1145/3506715} & 2022  & TIST  & CTR Prediction & Cross-platform \\
    FedCDR \cite{10.1145/3511808.3557320} & 2022  & CIKM 2022 & Rating prediction & Cross-platform \\
    FMSS \cite{10.1145/3548456} & 2022  & TOIS  & \multicolumn{1}{p{16.43em}}{Rating prediction,\newline{}Top-$k$,\newline{}Sequential recommendation} & Cross-client \\
    FedDSR \cite{9626622} & 2023  & TKDE  & Sequential recommendation & Cross-client \\
    PrivateRec \cite{liu2023privaterec} & 2023  & KDD 2023 & Top-$k$ recommendation & Cross-client \\
    FPPDM \cite{liufederatedP} & 2023  & IJCAI 2023 & Top-$k$ (cross-domain) & Cross-platform \\
    VFUCB \cite{cao2023privacy} & 2023  & KDD 2023 & CTR Prediction & Cross-platform \\
    SemiDFEGL \cite{10.1145/3543507.3583337} & 2023  & WWW 2023 & Top-$k$ recommendation & Cross-client \\
    ReFRS \cite{10.1145/3560486} & 2023  & TOIS  & Sequential recommendation & Cross-client \\
    PFedRS \cite{zhang2023dual} & 2023  & IJCAI 2023 & Top-$k$ recommendation & Cross-client \\
    LightFR \cite{zhang2023lightfr} & 2023  & TOIS  & Top-$k$ recommendation & Cross-client \\
    CPF-POI \cite{10241965} & 2024 & TKDE & POI recommendation & Cross-client \\
    FedCORE \cite{10443503} & 2024 & TKDE & Rating prediction & Cross-platform \\
    GPFedRec \cite{10.1145/3637528.3671702} & 2024 & KDD 2024 & Top-$k$ recommendation & Cross-client \\
    FedRAP \cite{li2024federated} & 2024 & ICLR 2024 & Top-$k$ recommendation & Cross-client \\
    CoFedRec \cite{10.1145/3589334.3645626} & 2024 & WWW 2024 & Top-$k$ recommendation & Cross-client \\
    FedHGNN \cite{10.1145/3589334.3645693} & 2024 & WWW 2024 & Top-$k$ recommendation & Cross-client \\
    PoisonFRS \cite{10.1145/3589334.3645492} & 2024 & WWW 2024 & Top-$k$ recommendation & Cross-client \\
    DGFedRS \cite{10.1145/3688570} & 2025 & TOIS & Top-$k$ recommendation & Cross-client \\
    PTF-FSR \cite{10.1145/3708344} & 2025 & TOIS & Top-$k$ recommendation & Cross-client \\
    FedDAE \cite{li2025personalized} & 2025 & AAAI 2025 & Top-$k$ recommendation & Cross-client \\

    \bottomrule
    \end{tabular}%
  \label{tab:FedRSs}%
\end{table}%

As mentioned before, traditional cloud-based recommender systems raise privacy concerns due to the fact that they collect all user data on a central server. Recently, inspired by the fact that federated learning has achieved promising performance in privacy-preserving machine learning, many research works have attempted to incorporate federated learning with recommender systems, termed federated recommender systems (FedRSs), to address privacy concerns. In the subsequent part, we will first provide a problem formulation of FedRSs, and then introduce the typical pipelines of FedRSs. Finally, we will categorize FedRSs based on the different methods adopted in the FedRS pipeline. Table \ref{tab:FedRSs} chronologically summarizes major FedRS methods and their addressed tasks, such as rating prediction \cite{chai2020secure,10.1145/3397271.3401081,lin2020fedrec,wu2022federated}, Top-$k$ recommendation \cite{10.1145/3394486.3403176,ammad2019federated,zhang2023lightfr,zhang2023dual}, and CTR prediction \cite{cao2023privacy,10.1145/3506715}. Additionally, some methods are designed to integrate specified side information in order to complete specific recommendation tasks. These include social recommendations that incorporate users’ social information \cite{chen2020secure,cui2021exploiting}, sequential recommendations based on user behavior information \cite{10.1145/3560486,10.1145/3548456}, and news recommendations integrating item attribute information \cite{qi2020privacy,liu2023privaterec}. 
Furthermore, FedRSs are categorized based on the types of participants. Cross-client FedRSs \cite{ammad2019federated,chai2020secure,10.1145/3548456} involve individual users or clients as training participants, each containing only their own user-item interaction data. In contrast, cross-platform FedRSs \cite{cui2021exploiting,chen2020secure,10.1145/3506715} involve recommendation service platforms or companies, each holding all user-item interaction data under their respective platform. The latter category often includes cross-domain recommendations \cite{10.1145/3404835.3462825,chen2022differential,liufederatedP}, which means the data sources from different platforms belong to various domains.


\subsubsection{Problem Formulation of FedRSs}
Let $\mathcal{U}$ and $\mathcal{I}$ denote a set of users and items, respectively. Each user $u \in \mathcal{U}$ holds a local dataset $\mathcal{D}_{u} = \{\mathcal{I}_{u},\textbf{y}_{u}\}$, where $\mathcal{I}_{u} \subseteq \mathcal{I}$ denotes a set of items interacted with user $u$, and $\textbf{y}_{u} \in \mathbb{R}^{|\mathcal{I}|}$ or $\textbf{y}_{u} \in \{0,1\}^{|\mathcal{I}|}$ represent explicit feedbacks (e.g., ratings) or implicit feedbacks (e.g., clicks and check-ins) of users, respectively. In addition, each device maintains a local model $f_{u}(\textbf{p}_{u},\textbf{Q}_{u},\Theta_{u})$
that typically consists of parameters including $d$-dimensional user embedding $\textbf{p}_{u} \in \mathbb{R}^{d}$, item embedding table $\textbf{Q}_{u} \in \mathbb{R}^{|\mathcal{I}| \times d}$, and model parameters $\Theta_{u}$. The server correspondingly maintains a global model $f_{s}(\textbf{Q}_{s},\Theta_{s})$ that typically does not include sensitive user embedding $\textbf{p}_{u}$. Finally, the goal of FedRSs is to learn the optimal global parameters $(\textbf{Q}_{s}^{*},\Theta_{s}^{*})$ by coordinating a set of devices to perform local training.


\subsubsection{Typical Pipeline of FedRSs}

\begin{figure*}
\setlength{\abovecaptionskip}{1pt}
    \centering
    \includegraphics[width=1\linewidth]{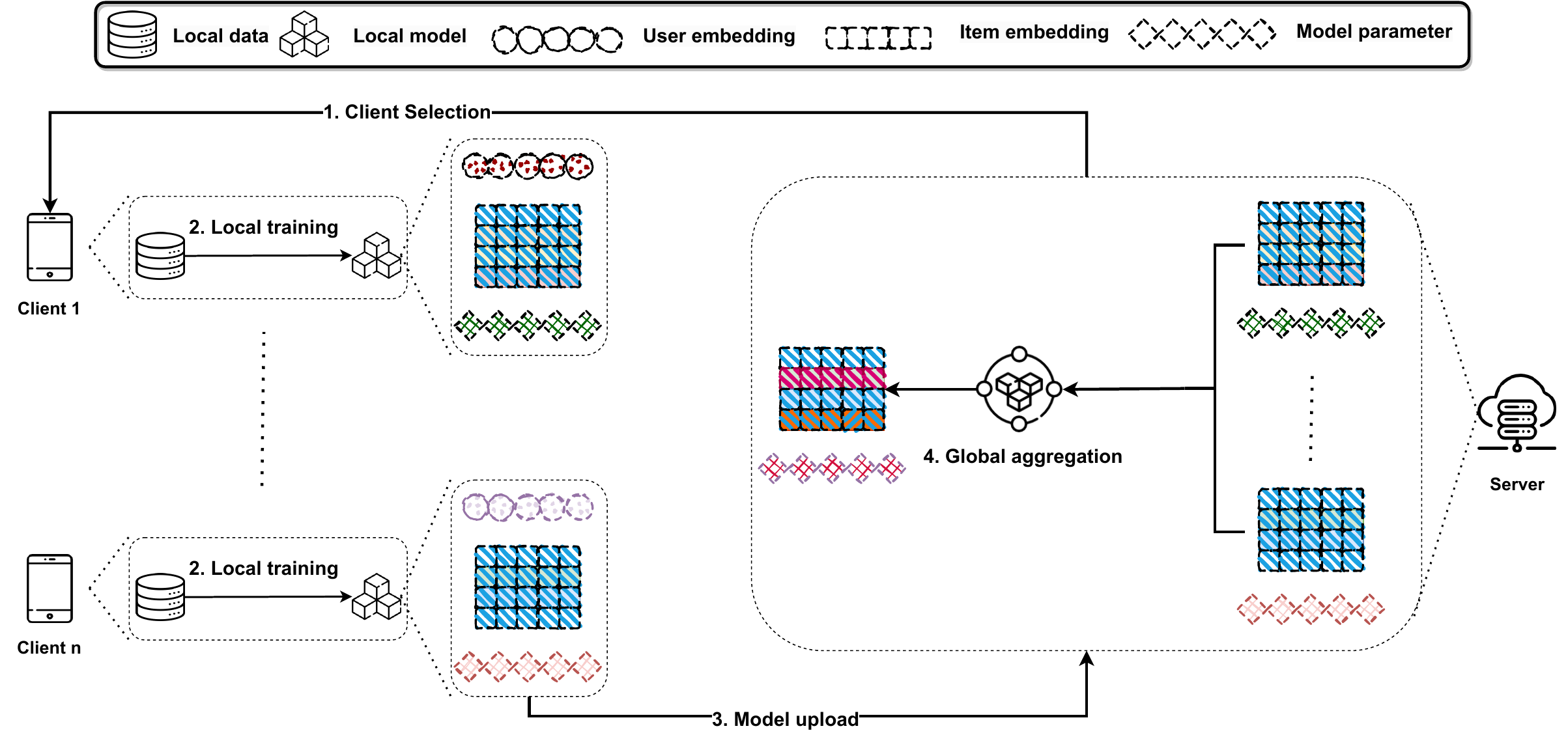}
    \caption{The pipeline of FedRSs typically consists of four steps including client selection, local training, model upload, and global aggregation.}
    \label{fig:FedRS}
\end{figure*}

As shown in Figure \ref{fig:FedRS}, the pipeline of FedRSs typically consists of four stages including: (1) \textbf{Client Selection:} for each training round, the server first selects a set of devices to participate in this round of training based on the predefined client selection strategy, and then distributes the global parameters of the server to these clients. (2) \textbf{Local Training}: the clients selected to participate in training perform training on a local model based on a local dataset. (3) \textbf{Model Ppload}: after local training, the device uploads parameters or corresponding gradients to the server. (4) \textbf{Global Aggregation:} the server employs the model aggregation strategy to obtain a global model based on uploaded parameters/gradients. After that, the updated global parameters are redistributed to each client. It is worth noting that there is no actual data sharing between devices and the server throughout the training pipeline of FedRSs for privacy concerns. Below, we will introduce the strategies used for the different methods in each stage, summarized in Table \ref{tab:FedRSspipeline}.

\begin{table}[htbp]
  \footnotesize 
  \centering
  \caption{Summary of the strategies used in the different stages of FedRSs.}
    \begin{tabular}{c|l|l}
    \toprule
    \multicolumn{1}{l|}{FedRS Pipeline} & \multicolumn{1}{c|}{Strategy} & \multicolumn{1}{c}{References} \\
    \midrule
    \midrule
    \multirow{3}[2]{*}{Client Selection} & Random selection & \multicolumn{1}{l}{\cite{10.1016/j.knosys.2022.108441,zhang2023dual,liu2023privaterec,10.1145/3548456,9626622,wu2022federated,10.1145/3511808.3557320,10.1145/3501815,liu2022fairness,10.1007/s00778-021-00700-6,10.1145/3404835.3462825,anelli2021federank,10.1145/3397271.3401081,qi2020privacy}} \\
          & Full selection & \multicolumn{1}{l}{\cite{ammad2019federated,10.1145/3442381.3449926,minto2021stronger,chai2020secure,lin2020fedrec,liang2021fedrec++,10.1145/3460231.3474257,liufederatedP,cao2023privacy,chen2022differential,10.1145/3506715,cui2021exploiting,chen2020secure,yang2021fcmf}} \\
          & Clustering-based selection & \cite{10.1145/3560486,10.1145/3543507.3583337,10.1145/3394486.3403176,10.1145/3511808.3557668} \\
    \midrule
    \multirow{4}[2]{*}{Local Training} & MF-FedRSs & \cite{ammad2019federated,chai2020secure,chen2020secure,anelli2021federank,cui2021exploiting,minto2021stronger,lin2020fedrec,liang2021fedrec++,10.1145/3442381.3449926,yang2021fcmf,liu2022fairness,zhang2023lightfr} \\
          & DNN-FedRSs & \cite{10.1145/3394486.3403176,qi2020privacy,10.1145/3397271.3401081,10.1145/3404835.3462825,10.1007/s00778-021-00700-6,chen2022differential,10.1016/j.knosys.2022.108441,10.1145/3506715,10.1145/3511808.3557320,10.1145/3548456,liu2023privaterec,10.1145/3560486,zhang2023dual}  \\
          & GNN-FedRSs: & \cite{wu2022federated,10.1145/3501815,10.1145/3511808.3557668,liufederatedP,10.1145/3543507.3583337}  \\
          & RL-FedRS: & \cite{10.1145/3460231.3474257,9626622,cao2023privacy} \\
    \midrule
    \multirow{7}[2]{*}{Model Upload} & Homomorphic Encryption  & \cite{chai2020secure,cui2021exploiting,yang2021fcmf,wu2022federated,10.1145/3560486}  \\
          & Secret Sharing    & \cite{10.1145/3460231.3478855,10.1145/3548456} \\
          & Secure Multiparty Computation   & \cite{chen2020secure,cui2021exploiting} \\
          & (Local) Differential Privacy & \cite{qi2020privacy,minto2021stronger,10.1007/s00778-021-00700-6,yang2021fcmf,chen2022differential,liu2022fairness,wu2022federated,10.1145/3501815,10.1145/3506715,liu2023privaterec,cao2023privacy,10.1145/3543507.3583337,10.1145/3560486,zhang2023dual} \\
          & Fake items    & \cite{lin2020fedrec,liang2021fedrec++,wu2022federated,10.1145/3501815,10.1145/3548456,10.1145/3543507.3583337} \\
          & Mask matrix     & \cite{10.1016/j.knosys.2022.108441} \\
          & Others & \cite{10.1145/3460231.3474257,10.1145/3442381.3449926,liufederatedP,zhang2023lightfr} \\
    \midrule
    \multirow{4}[2]{*}{Global Aggregation} & Gradient Descent & \cite{anelli2021federank,ammad2019federated,chai2020secure,10.1145/3397271.3401081,lin2020fedrec,liang2021fedrec++,10.1145/3460231.3474257,10.1145/3506715,cui2021exploiting,yang2021fcmf,chen2020secure}  \\
          & FedAvg & \cite{pmlr-v54-mcmahan17a,10.1145/3543507.3583337,10.1145/3548456,wu2022federated,10.1145/3511808.3557320,10.1145/3511808.3557668,10.1145/3501815,liu2022fairness} \\
          & Average Aggregation & \cite{zhang2023dual,liufederatedP,10.1007/s00778-021-00700-6,minto2021stronger,10.1145/3404835.3462825,qi2020privacy} \\
          & Others & \cite{10.1145/3442381.3449926,10.1016/j.knosys.2022.108441,liu2023privaterec,9626622,zhang2023lightfr,10.1145/3394486.3403176} \\
    \bottomrule
    \end{tabular}%
  \label{tab:FedRSspipeline}%
\end{table}%

\subsubsection{Client Selection}

In each training round, the server first selects a set of clients to participate in the current training round. It then distributes the latest global parameters to these clients. However, the data distribution of each client is usually non-independently and identically distributed (non-iid). Therefore, the method of selecting clients for each training round is crucial, as it can significantly affect the model's performance. In this section, we refer to this process as the client selection \cite{wang2022unified}, and introduce various client selection strategies below. 

\begin{itemize}
    \item \textbf{Random Selection} \cite{10.1016/j.knosys.2022.108441,zhang2023dual,liu2023privaterec,10.1145/3548456,9626622,wu2022federated,10.1145/3511808.3557320,10.1145/3501815,liu2022fairness,10.1007/s00778-021-00700-6,10.1145/3404835.3462825,anelli2021federank,10.1145/3397271.3401081,qi2020privacy}: Most existing FedRS methods adopt a random selection strategy, where a set of clients is selected randomly in each training round. Such client selection methods are generally applicable when there is a large number of clients, such as the cross-client-based FedRSs methods.

    \item \textbf{Full Selection} \cite{ammad2019federated,10.1145/3442381.3449926,minto2021stronger,chai2020secure,lin2020fedrec,liang2021fedrec++,10.1145/3460231.3474257}: Methods that select all clients for each training round are termed full selection. Although this type of methods require high communication costs between the server and clients, making them unsuitable for scenarios with a very large number of clients, they are generally applicable to cross-platform FedRS methods \cite{liufederatedP,cao2023privacy,chen2022differential,10.1145/3506715,cui2021exploiting,chen2020secure,yang2021fcmf} where each recommendation service platform serves as a client.

    \item \textbf{Clustering-based Selection} \cite{10.1145/3560486,10.1145/3543507.3583337,10.1145/3394486.3403176,10.1145/3511808.3557668}: There are also clustering-based methods for client selection. For example, ReFRS \cite{10.1145/3560486} calculates the semantic similarity of clients based on the model parameters they upload and employs dynamic clustering techniques to generate discriminative user clusters for later global aggregation. 
    SemiDFEGL \cite{10.1145/3543507.3583337} utilizes fuzzy c-means clustering to group the ego-graph embeddings uploaded by users into different clusters. Subsequently, clients are sampled individually within each group for global aggregation. PerFedRS \cite{10.1145/3511808.3557668} clusters users into a set of groups based on uploaded user embeddings, and within each group, clients are randomly selected in proportion to the cluster size. FedFast \cite{10.1145/3394486.3403176} introduces an active sampling strategy for client selection strategy, namely ActvSAMP, which utilizes k-means to cluster users into a set of groups based on their metadata or user embeddings and then randomly sample one client from each group to participate in federated training.
\end{itemize}

\subsubsection{Local Training}
After the client selection stage in each training round, each selected client performs local model training using the local dataset and global parameters received from the server. Specifically, each local dataset only contains the user's own interaction data between the user and items. In addition, the local parameters typically consist of three components including the user embedding, the item embedding table, and the model's parameters. In each training round, the global parameters received from the server are used to initialize the local model parameters. In this way, the client can perform standard model training locally. Below, we will introduce various recommendation methods utilized for local training, including matrix factorization-based methods (MF-FedRSs), deep neural networks-based methods (DNN-FedRSs), graph neural networks-based methods (GNN-FedRSs), and reinforcement learning-based methods (RL-FedRSs).

\begin{itemize}
    \item \textbf{MF-FedRSs} \cite{ammad2019federated,liu2022fairness,minto2021stronger,anelli2021federank,chai2020secure,lin2020fedrec,liang2021fedrec++,10.1145/3460231.3474257,yang2021fcmf} typically utilize the user embedding matrix $\textbf{P} \in \mathbb{R}^{|\mathcal{U}| \times d}$ and the item embedding $\textbf{Q} \in \mathbb{R}^{|\mathcal{I}| \times d}$ to represent users and items, respectively. Then, the preference $\textbf{R}$ of users $\mathcal{U}$ for items $\mathcal{I}$ can be predicted by $\textbf{R} = \textbf{P}  \textbf{Q}^{T}$.
    In the context of FedRSs, each user $u$ only maintain their own user embedding $\textbf{p}_{u}$ and item embedding table $\textbf{Q}_{u}$ locally. Formally, each client can calculate the local loss $\mathcal{L}_{u}$ based on the local dataset $\mathcal{D}_{u}$.
    The choice of loss function depends on the specific recommendation task, such as the Bayesian Personalized Ranking (BPR) in the Top-$k$ recommendation \cite{he2017neural}. In this way, the local user/item embeddings can be updated by the gradient descent locally.
    Furthermore, LightFR \cite{zhang2023lightfr} argues that the above real-valued MF methods could lead to privacy concerns and require substantial communication costs and storage space in the server-device communication process. To address this issue, it proposes a method based on a binary matrix by the learning-to-hash technique. Additionally, a corresponding discrete optimization method is introduced during the local training process to optimize the local binary matrix. HPFL \cite{10.1145/3442381.3449926} proposes a general user model (GUM) to perform local training, including public and private components. The former is used to learn non-sensitive information, such as item attributes, via multi-hot vectors, while the latter is employed to learn private user-item interaction information via MF. Finally, the knowledge matrix fuses the public item attribute representations with the user/item private embeddings, forming the final user/item representations. S$^{3}$Rec \cite{cui2021exploiting} and SeSoRec \cite{chen2020secure} leverage additional social information from another platform to alleviate the data sparsity issue in user-item rating matrix. They adopt Soreg \cite{ma2011recommender} to learn user/item embeddings based on both the user-item rating matrix and the user-user social matrix.

    \item \textbf{DNN-FedRSs} generally integrate additional deep neural network components with MF to capture the complex non-linear relationships between users and items. For example, some methods \cite{zhang2023dual,10.1145/3397271.3401081,10.1016/j.knosys.2022.108441} utilize Generalized Matrix Factorisation (GMF) \cite{he2017neuralcol} that employs multi-layer perceptron (MLP) as the score function to calculate the similarity score $s_{ij}$ between user embedding $\textbf{p}_{i}$ and item embedding $\textbf{q}_{j}$.
    Additionally, DNNs can also be used to learn user/item embeddings by leveraging side information such as the item attribute information. For instance, PrivateRec \cite{liu2023privaterec} utilizes an attention module and pre-trained models to learn news embeddings, and learns user embeddings based on the history of news clicked by the user. ReFRS \cite{10.1145/3560486} utilizes a self-supervised variational autoencoder (VAE) to learn user sequential interaction embeddings. FedCTR \cite{10.1145/3506715} proposes to use a neural user model to learn user embedding behaviors across different platforms in a hierarchical manner, thereby obtaining unified user embeddings to improve the performance of CTR prediction. This includes using a multi-head self-attention network to learn representations of user behaviors based on information such as search query behaviors and webpage titles, as well as employing an ad encoder model to learn the embeddings of advertisements.
    Moreover, there are hybrid methods that combine MF and DNNs to leverage both the linear representational power of MF and the non-linear relational capture ability of DNNs. For example, methods \cite{10.1145/3548456,10.1016/j.knosys.2022.108441} employ NeuMF \cite{he2017neuralcol} that ensembles MF and MLP to learn user/item embeddings. 
    Furthermore, DNN-FedRSs can also be applied to the federated cross-domain recommendation tasks. For example, FedCDR \cite{10.1145/3511808.3557320} utilizes the MLP as a transfer module to facilitate knowledge transfer in the cross-domain recommendation. PriCDR \cite{chen2022differential} employs the deep auto-encoder and DNN to learn embeddings on source-domain user-item interaction data and target-domain user-item interaction data, respectively. FedCT \cite{10.1145/3404835.3462825} introduces a modified VAE for both horizontal collaborations between clients and vertical collaborations between platforms for cross-domain recommendations. 

    \item \textbf{GNN-FedRSs} model user-item interactions as a user-item bipartite graph. Specifically, the data on each client can be viewed as a first-order user-centered ego graph, including only the user and the items they have directly interacted with. In this way, GNNs can be directly employed to learn user-item embeddings on each client. However, since the graphs on each local device are isolated first-order graphs and data cannot be shared with the server to construct a global graph due to privacy concerns, the full expressive power of GNNs cannot be utilized. To address this issue, SemiDFEGL \cite{10.1145/3543507.3583337} proposes a semi-decentralized federated recommendation framework, where, on the server side, clients are aggregated into different groups. In each group, fake common items are generated to connect the isolated ego graphs of each client, thereby establishing higher-order local subgraphs. In this way, existing GNN methods can be applied to these local subgraphs to learn higher-order graph structural information through device-to-device collaborative learning. FPPDM \cite{liufederatedP} employs federated learning to privacy-preserving cross-domain recommendation, i.e., each client is an independent recommendation service platform. Therefore, each client contains a global user-item bipartite graph so that the GNN can be directly utilized to learn the user/item embeddings. FedPerGNN \cite{wu2022federated} and PerFedRS \cite{10.1145/3511808.3557668} require an additional trusted third-party server to find and distribute anonymous neighbor information. This is done to expand the local ego graphs into a global graph. In this way, the server can collaborate with many local GNNs for learning. FeSoG \cite{10.1145/3501815} utilizes additional social information between users (i.e., user-user connections) to alleviate data sparsity and cold-start issues in the user-item bipartite graph. To this end, attention-based graph neural networks (e.g., GAT \cite{velivckovic2017graph}) are employed to learn heterogeneous relationship weights, such as user-item and user-user interactions, during neighbor aggregation.
    
    \item \textbf{RL-FedRSs} consider the recommendation problem as a sequential decision-making problem and employ reinforcement learning techniques to make recommendation decisions based on users' historical interaction data and user/item contexts. For example, FedDSR \cite{9626622} introduces integrating deep reinforcement learning with curriculum learning to perform daily schedule recommendations. VFUCB \cite{cao2023privacy} employs LinUCB \cite{abbasi2011improved} and LinTS \cite{agrawal2013thompson} to balance exploration and exploitation for online recommendations. To optimize the payload, FCF-BTS \cite{10.1145/3460231.3474257} adopts a classic reinforcement learning model, i.e., the multi-armed bandit, to intelligently select part of the item embeddings for distribution to all users resulting in reducing the communication costs between server and clients.
\end{itemize}

\subsubsection{Model Upload}

After the local training process, each participating client needs to upload locally updated parameters (e.g., $\textbf{Q}_{u}$) \cite{10.1145/3511808.3557320} or corresponding gradients (e.g., $\frac{\partial \mathcal{L}_{u}}{\partial \textbf{Q}_{u}}$) \cite{ammad2019federated,9626622,10.1145/3397271.3401081} to the server for the subsequent global model aggregation. Generally, parameters on each device consist of user embedding $\textbf{p}_{u}$, item embedding table $\textbf{Q}_{u}$, and model parameters $\Theta_{u}$. It is noteworthy that typically only $\textbf{Q}_{u}$ and $\Theta_{u}$ are uploaded to the server, while $\textbf{p}_{u}$ remains locally due to privacy considerations. For example, PFedRS \cite{zhang2023dual} proposes to only upload item embedding table $\textbf{Q}_{u}$ to the server, and the score function with a MLP parameterized by $\Theta_{u}$ is trained locally. In this way, each client can learn a personalized score function to capture user-specific preferences.

Although most of the current federated recommendation methods could enhance users' privacy due to raw data locality, the transmission of model updates, such as gradients or model parameters, from devices to the central server can still raise additional privacy issues. For example, in each training round, each device only updates the item embeddings corresponding to the items that have interacted with the user. So, the central server can easily recover the user's interaction history by comparing the difference between the newly updated item parameters uploaded by devices and the previous global item embedding table passed to devices \cite{8835269}. To mitigate this issue, FedRSs typically require some additional techniques to enhance privacy protections. For example, 
FPPDM \cite{liufederatedP} proposes to parameterize the locally trained embeddings with a Gaussian distribution to represent the user's preference and only uploads the obtained Gaussian distribution parameters, i.e., mean and covariance, to the global server. 
HPFL \cite{10.1145/3442381.3449926} proposes dividing local modal parameters into a non-sensitive public component, which can be directly uploaded to the server, and a sensitive private component. The private component is processed by performing a clustering task to obtain cluster centers as drafts representing user preferences, thereby reducing sensitivity. 
FMF-LDP \cite{minto2021stronger} introduces a proxy network to remove metadata and mix it with reports from other users, thereby breaking the linkability of consecutive reports from the same user.   
Furthermore, there are some widely used privacy protection techniques adopted in the model uploading process, including Homomorphic Encryption (HE), Secret Sharing (SS), Secure Multiparty Computation (MPC), Differential Privacy (DP), Local Differential Privacy (LDP), Fake items (FI), and Mask matrix (MM). We will elaborate these techniques in Section \ref{section:securityandprivacy}.

\subsubsection{Global Aggregation}
After the model upload process, the server aggregates these parameters/gradients to learn a global model. We refer to this step as the global aggregation. Below, we will introduce the various global aggregation strategies used by different methods.

\begin{itemize}
    \item \textbf{(Stochastic/Mini-batch) Gradient Descent} \cite{anelli2021federank,ammad2019federated,chai2020secure,10.1145/3397271.3401081,lin2020fedrec,liang2021fedrec++,10.1145/3460231.3474257,10.1145/3506715,cui2021exploiting}: Upon receiving gradients from users, one of the most straightforward methods is to use gradient descent to update global parameters with 
    $\textbf{Q}_{s} \leftarrow \textbf{Q}_{s} - \gamma \sum_{u \in \mathcal{U}^{+}} \frac{\partial \mathcal{L}_{u}}{\partial \textbf{Q}_{u}}$, 
    where $\mathcal{U}^{+}$ denotes the selected clients in this training round. The approach varies based on the value of $|\mathcal{U}^{+}|$: when $|\mathcal{U}^{+}|$=1, it functions as stochastic gradient descent (SGD) \cite{yang2021fcmf}; when $|\mathcal{U}^{+}|$=$|\mathcal{U}|$, it becomes gradient descent (GD); and when $1<|\mathcal{U}^{+}|$ < $|\mathcal{U}|$, it operates as a mini-batch gradient descent method \cite{chen2020secure}.

    \item \textbf{FedAvg}  \cite{pmlr-v54-mcmahan17a,10.1145/3543507.3583337,10.1145/3548456,wu2022federated,10.1145/3511808.3557320,10.1145/3511808.3557668,10.1145/3501815,liu2022fairness} is a widely used weighted average aggregation method. Taking the model parameter $\Theta_{u}$ aggregation as an example, FedAvg can be performed as $ \Theta_{s}^{(t+1)} \leftarrow \sum_{u \in \mathcal{U}^{+}}\frac{|\mathcal{D}_{u}|}{|\mathcal{D}|}\Theta_{u}^{(t)}$ for the current time step $t$,
    where $|\mathcal{D}_{u}|$ and $|\mathcal{D}|$ represent the size of the local dataset and the entire dataset, respectively. Additionally, there are other weighted aggregation methods that use different metrics as aggregation weights. For example, HPFL \cite{10.1145/3442381.3449926} employs the local model's validation accuracy as a dynamic aggregation weight.

    \item \textbf{Average Aggregation} \cite{zhang2023dual,liufederatedP,10.1007/s00778-021-00700-6,minto2021stronger,10.1145/3404835.3462825,qi2020privacy} calculates the average value of the parameters uploaded by all participating training clients in each round, which is then used as the global parameters as 
    $\Theta_{s}^{(t+1)} \leftarrow \sum_{u \in \mathcal{U}^{+}}\frac{1}{|\mathcal{U}^{+}|}\Theta_{u}^{(t)}$.

    \item \textbf{Other Methods:} Federated neural collaborative filtering (FedNCF) \cite{10.1016/j.knosys.2022.108441} introduces a novel aggregation method called MF-SecAvg, which extends the traditional FedAvg \cite{pmlr-v54-mcmahan17a} by integrating the secure aggregation (SecAvg) protocol \cite{10.1145/3133956.3133982} with a low computational cost. However, it can only protect the privacy of sampled clients in a single aggregation round. PrivateRec \cite{liu2023privaterec} employs FedAdam \cite{reddi2020adaptive} to update global model parameters based on uploaded gradients. ReFRS \cite{10.1145/3560486} proposes a user semantic-based global aggregation approach, aggregating only users with semantic similarities. As mentioned earlier, it first clusters users into different groups and then performs the aggregation in a group-wise manner to achieve quicker model convergence and better personalized recommendation performance. FedDSR \cite{9626622} proposes a similarity aggregation strategy that aggregates only similar clients. Specifically, it first calculates client-wise similarity based on the gradient parameters uploaded by the clients. Then, it selects clients for global aggregation based on a defined similarity threshold. LightFR \cite{zhang2023lightfr} proposes a tailored discrete aggregation method for binary matrices, achieved by applying the sign operation on the uploaded binary matrices. FedPerGNN \cite{wu2022federated} empirically explores several existing global aggregation methods, including FedAvg \cite{pmlr-v54-mcmahan17a}, FedAtt \cite{ji2019learning}, Per-FedAvg \cite{fallah2020personalized}, and pFedME \cite{t2020personalized}, and empirically finds that Per-FedAvg and pFedME can achieve better performance. FedFast \cite{10.1145/3394486.3403176} introduces an efficient federated aggregation strategy called ActcAGG, which involves ``delegates'' actively participating in training and sharing their progress with ``subordinates'' who are not involved in the training. This strategy updates non-embedding components similar to FedAvg, and item and user embeddings are updated based on weighted contributions from delegates, with a focus on actively updated components.

\end{itemize}


\subsection{Decentralized Recommender Systems}\label{section:DecRecs}

While FedRS partially alleviates users' concerns about personal data leakage, it presents significant limitations that warrant attention. A primary issue is its heavy reliance on a central server for collecting and aggregating user models, as well as redistributing the combined model. This dependence demands substantial computational and storage resources from the server and exerts considerable strain on network bandwidth, leading to elevated economic and environmental costs. Another critical drawback is the `long-tail problem' arising from all users in FedRS sharing the same model. This results in the aggregated model's inability to effectively capture diverse user preferences. 

In response, decentralized recommender systems (DecRSs) have been developed to overcome the limitations of FedRSs. Unlike standard FedRSs, DecRSs optimize on-device models through a combination of local training and inter-device communication within specific user groups. This approach significantly minimizes central server involvement, thereby reducing privacy risks and enabling fully personalized recommendation services for users. In the subsequent sections, we will formally define DecRSs and outline their operational pipeline. We will also discuss various DecRSs, differentiated by the strategies employed within their pipeline, as summarized in Table \ref{tab:DecRSspipeline}.

\subsubsection{Problem Formulation of DecRSs}

Let $\mathcal{U}$ and $\mathcal{I}$ denote a set of users and items, respectively. Each user $u \in \mathcal{U}$ holds a local dataset $\mathcal{D}_{u} = \{\mathcal{I}_{u},\textbf{y}_{u}\}$, where $\mathcal{I}_{u} \subseteq \mathcal{I}$ denotes a set of items interacted with user $u$, and $\textbf{y}_{u} \in \mathbb{R}^{|\mathcal{I}|}$ or $\textbf{y}_{u} \in \{0,1\}^{|\mathcal{I}|}$ represent explicit feedbacks (e.g., ratings) or implicit feedbacks (e.g., clicks and check-ins) of users, respectively. In addition, each device maintains a local model $f_{u}(\textbf{p}_{u},\textbf{Q}_{u},\Theta_{u})$
that typically consists of parameters including $d$-dimensional user embedding $\textbf{p}_{u} \in \mathbb{R}^{d}$, item embedding table $\textbf{Q}_{u} \in \mathbb{R}^{|\mathcal{I}| \times d}$, and model parameters $\Theta_{u}$. More importantly, each user can obtain extra knowledge $\mathcal{K}_{u}$ from the corresponding neighbors $\mathcal{N}_{u}$. Given these, the goal of DecRSs is to train a local model for each user by minimizing loss functions on the device side $\mathcal{L}_{u}(\mathcal{D}_{u},f_{u}, \mathcal{K}_{u})$.

\subsubsection{Typical Pipeline of DecRSs}

\begin{figure*}
\setlength{\abovecaptionskip}{1pt}
    \centering
    \includegraphics[width=1\linewidth]{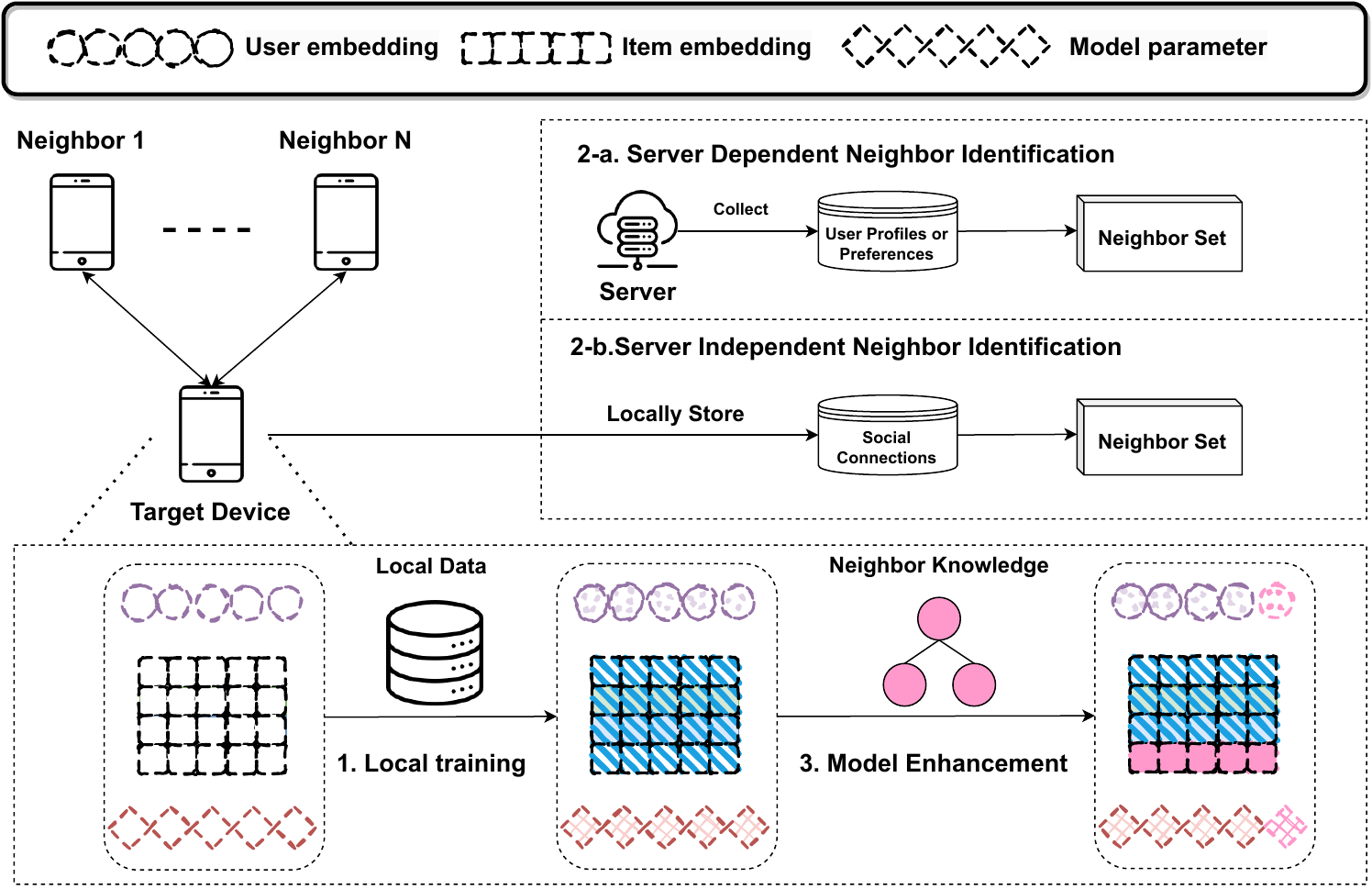}
    \caption{The pipeline of DecRSs.}
    \label{fig:DecRec}
\end{figure*}

As shown in Figure~\ref{fig:DecRec}, the implementation of DecRSs can be divided into three stages, including (1) Neighbor Identification: for each user, neighbors for collaborative learning are identified with predefined neighbor identification strategies before the training process or at the start of each training round. (2) Local Training: each user trains a local model with personal data. (3) Collaborative Learning with Neighbors: the locally trained model is further enhanced by exchanging knowledge with neighbors. Strategies used in each stage will be further discussed in the following sections.

\begin{table}[htbp]
  \small 
  \centering
  \caption{Summary of the strategies used in the different stages of DecRSs.}
    \begin{tabular}{c|l|l}
    \toprule
    \multicolumn{1}{l|}{DecRS Pipeline} & \multicolumn{1}{c|}{Strategy} & \multicolumn{1}{c}{References} \\
    \midrule
    \midrule
    \multirow{2}[2]{*}{Local Training} 
          & MF &  \cite{wang2015DRS,kermarrec2010RW,chen2018DMF,ling2012DLMC,hegedHus2019decentralized,defiebre2020decentralized,defiebre2022human,yang2022dpmf,an2024nrdl}\\
          & GNN &  \cite{zheng2023decentralized,10.1145/3543507.3583337}\\
          & DNN &  \cite{long2023decentralized,long2023model}\\
    \midrule
    \multirow{11}[2]{*}{Neighbor Identification} 
          & Social Network-based &  \cite{wang2015DRS, yang2022dpmf, long2023model, an2024nrdl}\\
          & Random Assignment &   \cite{kermarrec2010RW, ling2012DLMC, hegedHus2019decentralized, zheng2023decentralized}\\
          & General Profile-based &  \cite{defiebre2020decentralized,defiebre2022human}\\
          & Geographical Information-based &  \cite{chen2018DMF, beierle2019collaborating, long2023decentralized,long2023model}\\
          & User Embedding-based &  \cite{defiebre2020decentralized,defiebre2022human,10.1145/3543507.3583337}\\
          & Item Embedding-based &  \cite{defiebre2020decentralized,defiebre2022human}\\
          & Response-based &  \cite{long2023model}\\
          & Category Distribution-based &  \cite{long2023decentralized, long2023model}\\
          & Random Walk &  \cite{kermarrec2010RW,chen2018DMF,an2024nrdl}\\
          & Two-Step Neighbor Identification &  \cite{kermarrec2010RW, long2023model}\\
          & Performance-triggered Sampling &  \cite{long2023model}\\
    \midrule
    \multirow{4}[2]{*}{Model Upload} 
          & Data Sharing  &   \cite{kermarrec2010RW,defiebre2020decentralized,defiebre2022human,beierle2019collaborating}\\
          & Model Sharing  &  \cite{wang2015DRS,ling2012DLMC,hegedHus2019decentralized,10.1145/3543507.3583337,long2023decentralized}\\
          & Gradient Sharing  &  \cite{chen2018DMF,yang2022dpmf,an2024nrdl,zheng2023decentralized}\\
          & Soft Decision Sharing &  \cite{long2023model}\\
    \bottomrule
    \end{tabular}%
  \label{tab:DecRSspipeline}%
\end{table}%

\subsubsection{Local Training}

To facilitate neighbor identification and collaborative learning, each user initially trains a local model using their personal data. Echoing the methodology of FedRSs, the recommendation strategies implemented at the user level predominantly encompass Matrix Factorization (MF), Graph Neural Networks (GNNs), and Decurrent Neural Networks (DNNs). The primary aim of these approaches is to derive insightful user and item embeddings that accurately represent their characteristics and preferences. To prevent redundancy in our discussion, we only introduce two distinct loss functions tailored for separate recommendation tasks here: one for rating prediction and the other for click-through rate optimization.

\textbf{Mean Square Error (MSE).} For user-item ratings, user and item embeddings are updated using the Mean Square Error (MSE), defined as 
$\mathcal{L}_{MSE} = \sum_{r_{mn}\in R}(r_{mn}-u_{m}^{T}i_{n})^2$
where $ u_{m} $ and $ i_{n} $ represent the embeddings of user $ m $ and item $ n $ respectively. $ r_{mn} $ denotes the rating given by user $ u_m $ for item $ i_n $, and $ R $ encompasses all known user-item ratings. 

\textbf{Cross-Entropy Loss (CE).} In many real-world scenarios, explicit feedback, such as ratings, is not always available. Instead, data often comprises implicit feedback, reflecting user-item interactions. In these cases,CE loss is employed. The CE loss is formulated as:
$\mathcal{L}_{CE} = -\sum_{y_{mn}\in Y^{+} \cup Y^{-}}\left(y_{mn}\log(\hat{y}_{mn}) + (1-y_{mn})\log(1-\hat{y}_{mn})\right)$
where $ Y^{+} $ denotes the set of positive user-item interactions and $ Y^{-} $ represents an equivalent set of randomly sampled negative interactions. In this context, $ y_{mn} = 1 $ indicates an interaction between user $ u_m $ and item $ i_n $, while $ y_{mn} = 0 $ denotes no interaction. The predicted interaction probability $ \hat{y}_{mn} $ is calculated as:
$\hat{y}_{mn} = \text{sigmoid}(u_{m}^{T}i_{n})$.

\subsubsection{Neighbor Identification}

Training a recommender system solely on isolated devices faces a severe challenge of data sparsity. Meanwhile, communicating with all other devices poses privacy concerns, high communication costs, and long-tail issues. Hence, DecRSs aim to identify quality neighbors for collaborative learning. Among existing works of DecRSs, neighbor identification approaches can be broadly categorized into two types: (1) \textbf{Server-Independent Approaches} and (2) \textbf{Server-Dependent Approaches}. This section will comprehensively present various methods under the two categories.

\textbf{Server-Independent Approaches}. In these approaches, neighbor identification is entirely conducted on the user's side. In DecRSs, social information is stored locally, enabling users to interact with their friends. Given that friends often share high preference affinities, considering them as potential high-quality neighbors is a sensible strategy \cite{wang2015DRS, yang2022dpmf, long2023model, an2024nrdl}. To expand the scope of communication, users may also connect and exchange information with friends of friends, provided all parties consent \cite{an2024nrdl}. However, relying solely on local social networks for neighbor discovery can be limiting. Quality neighbors with similar preferences may not be part of one's immediate social circle. For instance, users with a shared interest in a specific anime genre are unlikely to be confined to a single social network. Therefore, recent research has shifted towards identifying high-quality neighbors globally.

\textbf{Server-Dependent Approaches}. Under this category, a server is utilized to construct a user adjacency matrix $ W $, where each element $ W_{a,b} \in [0,1] $ indicates the correlation degree between users $ u_a $ and $ u_b $. Based on this, the $ N $ users most correlated with $ u_a $ are chosen as their neighbors, denoted as $ \mathcal{N}(u_a) $. The primary challenge lies in populating this matrix. One basic method is \textbf{random assignment}, leading to $ \mathcal{N}(u_a) $ being a random subset of the user set $ \mathcal{U} $ \cite{kermarrec2010RW, ling2012DLMC, hegedHus2019decentralized, zheng2023decentralized}. It's important to note that this method cannot be implemented locally without access to all device IDs. Beyond random assignment, user correlation is typically quantified based on similarities in profiles and preferences. The specific factors contributing to this correlation will be detailed subsequently:
\begin{itemize}

    \item \textbf{General Profile.} The fundamental components of a user profile usually encompass age, gender, education level, and preference level for various item categories. When users exhibit similar profiles, it implies a shared commonality in their preferences, making them potentially valuable neighbors in recommendation systems. As outlined in \cite{defiebre2020decentralized,defiebre2022human}, the similarity between the profile vectors \( p_{u_a} \) and \( p_{u_b} \) of two users is quantified as
    $s_{p}(u_a,u_b) = \frac{1}{1 + d(p_{u_a}, p_{u_b})}$,
    where $ d(\cdot) $ represents a chosen distance function, which could be the Euclidean distance, Manhattan distance, or Cosine distance.

    \item \textbf{Geographical Information.} As a distinctive aspect of user profiles, geographical information plays a pivotal role in categorizing users for recommendation systems. This is due to its significant impact on users' preferences for certain products and services, which can be influenced by factors like climate, transportation, and local culture, all closely tied to geographical location. Such influences are particularly evident in Point-Of-Interest (POI) recommendations, where only users within the same region have access to the same POIs \cite{long2023model}. In light of this, the geographical similarity $ s_{g}(u_a,u_b) $ between two users $ u_a $ and $ u_b $ is defined as $ 1 $ if they are in the same region, and $ 0 $ otherwise. Regions can be determined based on administrative boundaries, with adjustable granularity. Furthermore, references \cite{chen2018DMF, beierle2019collaborating, long2023decentralized} propose an alternative quantification of geographical similarity as
    $s_{g}(u_a,u_b) = \frac{1}{1 + d_{geo}(u_a, u_b)}$,
    where $ d_{geo}(u_a, u_b) $ denotes the physical distance between the users $ u_a $ and $ u_b $.

    \item \textbf{User Embedding}. As previously mentioned, each user initially learns user and item embeddings using locally stored data before engaging in collaborative learning with neighbors. These user embeddings, which encapsulate individual preferences, can be utilized to calculate user similarity  \cite{defiebre2020decentralized,defiebre2022human} as
    $s_{u}(u_a,u_b) = \frac{1}{1 + d(e_{u_a}, e_{u_b})}$, 
    where $e_{u_a}$ and $e_{u_b}$ represent the embeddings of users $u_a$ and $u_b$ respectively, and $d(\cdot)$ is a distance function, such as Euclidean distance, Manhattan distance, or Cosine Distance. Beyond simple pairwise similarity calculations, \cite{10.1145/3543507.3583337} employs fuzzy c-means clustering to group users based on their embeddings. In this framework, $s_{u}(u_a,u_b)$ is assigned a value of 1 if $u_a$ and $u_b$ belong to the same cluster and 0 otherwise.

    \item \textbf{Item Embedding}. Similar to user embeddings, users with similar item embeddings are also considered similar \cite{defiebre2020decentralized,defiebre2022human}. The item similarity is formally defined as
    $s_{i}(u_a,u_b) = \sum_{i \in I_{u_a} \cap I_{u_b}} \frac{e_{i}^{u_a} \cdot e_{i}^{u_b}}{||e_{i}^{u_a}|| \times ||e_{i}^{u_b}||}$,
    where $I_{u_a}$ and $I_{u_b}$ denote the sets of items interacted with by users $u_a$ and $u_b$, respectively. It is worth noting that the item similarity is not averaged by the number of overlapping items because the greater number of overlapping items inherently implies higher user similarity.

    \item \textbf{Response on Reference Dataset}. To mitigate privacy concerns associated with uploading personal embeddings, MAC \cite{long2023model} proposes using soft decisions derived from applying the personal model $\Theta$ to a public reference dataset $\mathcal{D}^{ref}$. This approach provides a less sensitive yet effective representation of user preferences. User similarity in this context is quantified using the Kullback-Leibler (KL) divergence as
    $s_{r}(u_a,u_b) = \sum_{\forall \mathcal{X} \in \mathcal{D}^{ref}} KL(\Theta_{a}(\mathcal{X}) || \Theta_{b}(\mathcal{X}))$.

    \item \textbf{Category Distribution}. In recommendation systems, the prevalent issue of data sparsity often makes it difficult to identify a sufficient number of high-quality neighbors based solely on overlapping items. An alternative approach is to consider the categories of items interacted with by users. These category distributions, being denser and less sensitive than individual items, offer a secure way to reflect similarities in user preferences. In this context, \cite{long2023decentralized, long2023model} employ the Kullback-Leibler (KL) divergence to measure the category similarity as
    $s_{c}(u_a,u_b) = KL(CP(u_a) || CP(u_b))$,
    where $CP(u_a)$ and $CP(u_b)$ represent the distributions across all item categories for users $u_a$ and $u_b$, respectively.

\end{itemize}

After collecting the corresponding information, the server can build the user adjacency matrix $W$ by calculating and normalizing user similarities. In addition, a \textbf{random walk approach} can be further utilized to refine the matrix \cite{kermarrec2010RW,chen2018DMF,an2024nrdl}. This method capitalizes on the principle of transitive correlations among users: if $u_a$ and $u_b$ exhibit a high degree of correlation, and similarly, $u_b$ and $u_c$ are closely correlated, it is reasonable to infer a significant correlation between $u_a$ and $u_c$. The random walk algorithm navigates through these interconnected user relationships, enabling the system to identify and reinforce these indirect yet meaningful correlations. Consequently, the adjacency matrix becomes more robust, accurately reflecting the nuanced network of user correlations, both direct and indirect.

While proven effective in identifying neighbors, existing methods also present notable drawbacks, broadly falling into two categories. Firstly, the quality of neighbors identified by some methods varies, leading to less effective collaborative learning. A prime example is the random-based method, which often pairs users with neighbors having little relevance to their interests. Moreover, even if users share similar profiles, geographical locations, or category distributions, their product preferences may not fully align. Consequently, not all neighbors identified by these methods are genuinely meaningful. Secondly, some methods that can accurately define high-quality neighbors necessitate users to share sensitive information (such as personal embeddings or even soft decisions) with the server, posing significant privacy concerns. In summary, there remains an unmet need for a method capable of identifying high-quality neighbors without compromising user privacy.

To bridge this gap, a \textbf{two-step neighbor identification} method is proposed \cite{kermarrec2010RW, long2023model}. Initially, a server utilizes non-sensitive information (e.g., de-identified geographical information and category distributions) to identify a preliminary set of neighbors. The user device then refines this set using more sensitive yet representative data. This approach enables users to obtain high-quality neighbors for collaborative learning while minimizing privacy risks. In addition, the fluctuations in the local loss function are valuable indicators of the current neighbors' informativeness. Thus, \cite{long2023model} introduces a \textbf{performance-triggered sampling} strategy. A small subset of neighbors is initially sampled from the pre-selected pool. During each collaborative learning (CL) epoch $o$, if the change in $\Delta L_{loc}^o(u_a)$ is less than a predefined threshold $\tau = 1\%$, the neighbors are reselected from the pre-selected set. The change in local loss, $\Delta L_{loc}(u_i)$, is defined as
$\Delta L_{loc}^o(u_i) = \frac{|L^{o}_{loc}(u_i) - L^{o-1}_{loc}(u_i)|}{L^{o-1}_{loc}(u_i)} \times 100\%$,
where the $\tau$ is adjustable and controls the sensitivity to performance changes.

\subsubsection{Collaborative Learning with Neighbor}
Given the knowledge from high-quality neighbors $\mathcal{N}_a$ and the corresponding similarities, the next step is to enhance the personalized model $\Theta_a$ with such information. According to different knowledge shared between users, there are currently four types of collaborative learning approaches: data sharing approach, model sharing approach, gradient sharing approach, and soft decision sharing approach. To avoid repetitive narration, the common symbols used in this section are listed here: $\gamma$ is the learning rate, $\mu$ controls the strength of the neighbor knowledge, $\mathcal{W}_{ab}$ represents the similarity of $u_a$ and $u_b$.

\textbf{Data Sharing Approach}. The straightforward approach is to share data (i.e., ratings or interacted items) with neighbors for collaborative learning \cite{kermarrec2010RW,defiebre2020decentralized,defiebre2022human,beierle2019collaborating}. Upon receiving the data from the neighbors, the personal model can be further updated with SGD, formulated as
$\Theta_{a} \leftarrow \Theta_{a} - \gamma \mu \frac{1}{|\mathcal{N}_a|}\sum_{u_b \in \mathcal{N}_a} \mathcal{W}_{ab}\frac{\partial \mathcal{L}(\Theta_a,\mathcal{D}_b)}{\partial \Theta_{a}}$,
where $\mathcal{L}(\cdot)$ is either $\mathcal{L}_{MSE}$ for ratings or $\mathcal{L}_{CE}$ for interacted items, and $\mathcal{D}_b$ refers to personal data of $u_b$.

\textbf{Model Sharing Approach}. Even within a small scope, directly sharing private data carries significant risks of privacy breaches. On this basis, recent studies of DecRSs allow users to share models, keeping private data on the user side consistently \cite{wang2015DRS,ling2012DLMC,hegedHus2019decentralized,10.1145/3543507.3583337,long2023decentralized}. Given the neighbor models, the enhanced model is defined as
$\Theta_a \leftarrow \Theta_a + \mu\frac{1}{|\mathcal{N}_a|}\sum_{u_b\in\mathcal{N}_a}\mathcal{W}_{ab}\Theta_b$.

\textbf{Gradient Sharing Approach}. To further mitigate the risk of privacy leakage, gradients are utilized as the bridge for collaborative learning instead of the model \cite{chen2018DMF,yang2022dpmf,an2024nrdl,zheng2023decentralized}. Then, the personal model enhanced by such information is formulated as
$\Theta_{a} \leftarrow \Theta_{a} - \gamma \mu \frac{1}{|\mathcal{N}_a|}\sum_{u_b \in \mathcal{N}_a} \mathcal{W}_{ab}\frac{\partial \mathcal{L}_{b}}{\partial \Theta_{b}}$.
However, some methods \cite{zhu2019deep,yin2021see} successfully reconstructed the model with all gradients, proving that sharing models and sharing gradients possess nearly the same level of privacy security and collaborative effectiveness.

\textbf{Soft Decision Sharing Approach}. Unfortunately, frequently transporting models or gradients breaks the purpose of communication efficiency. In addition, both approaches have a strong assumption that all users have homogeneous models. This assumption significantly harms the practicality of DecRSs as each local model should have a customized structure to meet the specific device. In light of this, MAC \cite{long2023model} proposes to let $u_a$ get close to her neighbors by minimizing their disagreement of soft decisions on the public reference dataset $\mathcal{D}^{ref}$, which is quantified as
$L_{SD} = \frac{1}{|\mathcal{N}_a|}\sum_{u_b\in \mathcal{N}_a}\mathcal{W}_{ab} \bigg{(} \sum_{\mathcal{X} \in \mathcal{D}^{ref}}\left|\left|\Theta_a
  \left(\mathcal{X}\right)-\Theta_b(\mathcal{X})\right|\right|^2_2 \bigg{)}$. Then, the enhanced model is defined as
$\Theta_{a} \leftarrow \Theta_{a} - \gamma \mu \frac{\partial \mathcal{L}_{SD}}{\partial \Theta_{a}}$.


\begin{figure*}
\setlength{\abovecaptionskip}{1pt}
    \centering
    \includegraphics[width=\linewidth]{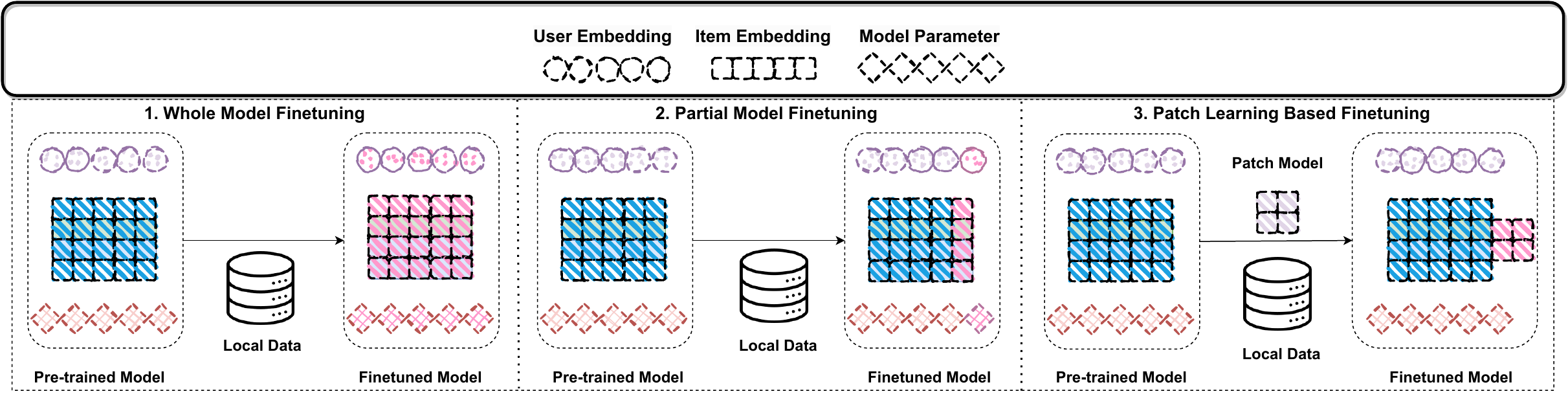}
    \caption{Three types of on-device finetuning methods.}
    \label{fig:FTRec}
\end{figure*}

\subsection{On-device Recommender Finetuning}\label{section:DeviceRSfinetuning}

Although device-cloud collaborative learning (FedRSs) and device-device collaborative learning (DecRSs) have demonstrated their ability to provide accurate yet secure recommendations immediately, they both require users to have sufficient computation resources to participate in the training process, impairing their practicality. Moreover, the extensive training time might also annoy inactive users. To solve this problem, researchers proposed another on-device learning approach known as on-device recommender finetuning. Specifically, a server is initially involved in training a global model and distributing the global model to all user devices. Then, this model is adapted to each user by fine-tuning with local data. Since the deployment approach is well-discussed in existing studies, which are discussed in Section \ref{section:deployandinference}, researchers in this domain focus on making the deployed model adapt to user data quickly and effectively. In summary, three types of fine-tuning methods will be discussed in this section and demonstrated in Figure \ref{fig:FTRec}. To reduce redundancy, the common notations in this section are listed here: $\gamma$ is the learning rate, $\mathcal{L}(\cdot)$ is either $\mathcal{L}_{MSE}$ for ratings or $\mathcal{L}_{CE}$ for interacted items, and $D_a$ is the local data of $u_a$.

\textbf{Whole Model Finetuning.} The most straightforward way is to update the whole network with the local data \cite{10.1007/s00778-021-00700-6,nichol2018first,yan2022device}. Formally, given the deployed model $\Theta_s$ pre-trained on the server, the personalized model on the device side for $u_a$ is defined as
$\Theta_{a} \leftarrow \Theta_{s} - \gamma \frac{\partial \mathcal{L}(\Theta_s,D_a)}{\partial \Theta_{s}}$.
However, given the extensive scale parameters involved in the cloud model, adapting the entire network would be an impractical approach.

\textbf{Partial Model Finetuning.} Alternatively, method \cite{mairittha2020improving} proposes to finetune partial parameters. That is, after $u_a$ receives the pre-trained model $\Theta_s$, each item embedding can be decomposed into two parts:
$e_i = e^s_i + e^a_i$,
where $e^s_i$ represents stable parameters, while $e^a_i$ denotes the variable part and its size can be adjusted. Then, $e^a_i$ is updated with the local data as
$e^a_i \leftarrow e^a_i - \gamma \frac{\partial \mathcal{L}(\Theta_s,D_a)}{\partial e^a_i}$.
Unfortunately, only finetuning partial parameters is performance-limited as the awkward combination of stable and variable vectors disrupted the original features.

\textbf{Patch Learning-based Finetuning.} Hence, patch learning is introduced to address the shortcomings of partial model finetuning, which has been proven to achieve comparable performance with whole model finetuning \cite{yao2021device}. Concretely, after deploying the pre-trained model $\Theta_s$ on the device of $u_a$, a personalized patch model $\Psi_a$ is inserted and the adjusted model is defined as
$\Theta_a(\cdot) \leftarrow \Theta_s(\cdot) + \mu \Psi_a(\Theta_s(\cdot))$.
where $\mu$ controls the level of the adaptation effect. Then, only the patch model is uploaded with the local data by
$\Psi_a \leftarrow \Psi_a - \gamma \frac{\partial \mathcal{L}(\Theta_a,D_a)}{\partial \Psi_a}$.
Note that, the patch model could have different neural architectures if its inputs and outputs are aligned with $\Theta_s$.

\section{Security for DeviceRSs}\label{section:securityandprivacy}
With the notable accomplishments that DeviceRSs have achieved in both academia and industry in recent years, numerous researchers have embarked on exploring DeviceRSs from a security standpoint. 
Generally, security studies related to DeviceRSs encompass two crucial aspects: safeguarding privacy and ensuring the robustness of systems. 
In the subsequent subsections, we present a thorough literature review to introduce security research undertaken in these two dimensions.

\subsection{Privacy Risks and Countermeasures}\label{section:privacyrisksandcountermeasures}
\begin{figure}
\setlength{\abovecaptionskip}{1pt}
  \centering
  \includegraphics[width=1\linewidth]{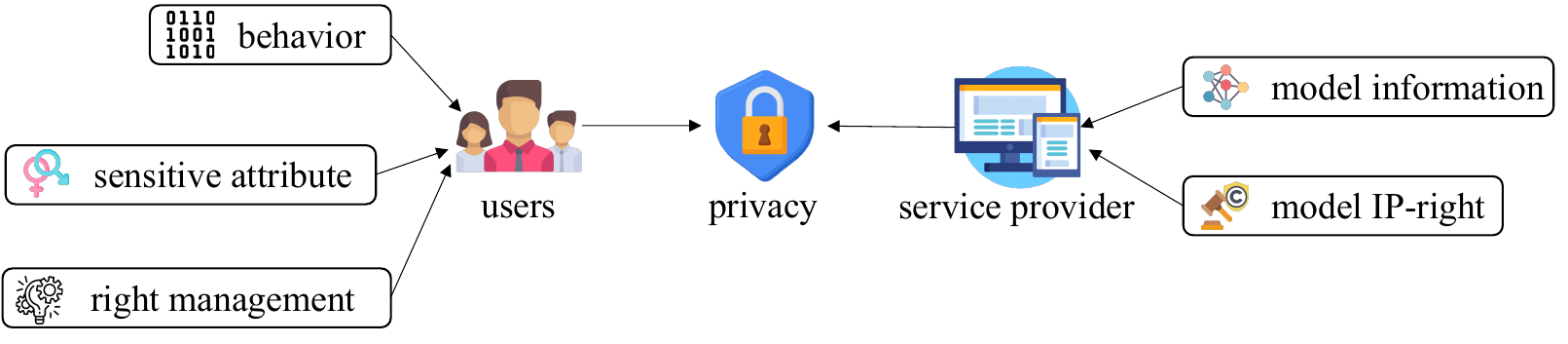}
  \caption{Different perspectives of privacy in DeviceRSs.}
  \label{fig_privacy_perspectives}
\end{figure}

Within the context of DeviceRSs, there exist two primary participant categories: users and service providers. 
These distinct participant types harbor entirely different privacy requirements. Figure~\ref{fig_privacy_perspectives} delineates the privacy considerations in DeviceRSs as perceived by these diverse participants. 
For users, paramount privacy concerns revolve around the protection of their behavioral data, sensitive information, and data management rights. 
In contrast, service providers place their privacy emphasis predominantly on model information and model intellectual property rights. 
In the ensuing discussion, we will delve into privacy considerations within DeviceRSs from both the users' and service providers' perspectives.

\subsubsection{Privacy Issues and Solutions from User Perspectives}\label{sec_user_privacy}

Compared to CloudRSs, DeviceRSs shift the training process to the devices' side considering user privacy. For example, in a typical DeviceRS (such as a federated recommender system), users' raw data are consistently stored on their respective local devices and remain inaccessible to other participants, and the model is trained based on local data while aggregate parameters with other participants to achieve collaborative learning. 
Despite the privacy-centric design of DeviceRSs, numerous studies have delved into and verified their actual privacy protection capabilities. 
These investigations have revealed that, in practice, DeviceRSs encounter several challenges related to user privacy.

\begin{itemize}
  \item \textbf{User Behavior Data Leakage.} 
  User behavior data encompass both explicit behaviors, such as user review ratings, and implicit information, including clicking, browsing, and purchasing. 
  When recommending based on explicit feedback, where uploaded item embedding gradients pertain solely to users' rated items, a central server with inquisitive intentions can easily identify rated items by analyzing non-zero gradients~\cite{wu2021fedgnn}. 
  Furthermore, FedMF~\cite{chai2020secure} indicates that by scrutinizing gradients sent by clients over two consecutive rounds, the central server can even infer the rating scores of items, leading to compromised rating information within the current mechanism.
  For recommendation systems using implicit feedback, all non-interacted items are treated as negative samples, and the uploaded item gradients encompass both user-interacted and non-interacted items. 
  Consequently, these methods appear to offer enhanced privacy protection. However, Yuan et al.~\cite{yuan2023interaction} reveal that a curious but honest central server can accurately deduce users' interacted item sets by calculating the similarity between mimic item embeddings trained on a guessing dataset and the original uploaded item embeddings.
  \item \textbf{User Sensitive Attribute Leakage.}
  Except for user behavior information, customers in recommender systems have other personal attributes that need to be protected, including but not limited to gender, age, occupation, political orientation, health status, etc.~\cite{zhang2021graph}.
  In the context of DeviceRSs, where the interaction data is maintained by clients and the central server lacks access to user recommendation lists, conventional attribute inference attack methods designed for centralized recommender systems are inapplicable~\cite{beigi2020privacy, zhang2021graph}.
  The pioneering work of~\cite{zhang2023comprehensive} delves into the potential threats of attribute leakage in DeviceRSs. 
  Their research highlights that in the presence of a small proportion of data leakages, a curious-but-honest central server can train an attacker with a minimal number of Multilayer Perceptron (MLP) layers to achieve high inference accuracy. 
  \item \textbf{Users' Data Ownership.}
  Most recent privacy regulations~\cite{harding2019understanding} have claimed that users have the right to control and withdraw their data contributions at any time, named ``The Right to Be Forgotten''.
  Unfortunately, the work~\cite{yuan2023federated} suggests that there is a lack of consideration for effectively removing the contributions of the user.
  In light of this, FRU~\cite{yuan2023federated} employs a federated unlearning method to efficiently remove specific clients' influence via historical update calibration.
\end{itemize}

As shown in the above description, although DeviceRSs offer certain privacy protection for users, they still have many privacy issues.
To further improve the privacy-preserving ability, many works incorporate various privacy protection mechanisms in DeviceRSs. These methods can generally be classified into two types: obfuscation-based approaches and encryption-based approaches.

\textbf{Obfuscation-based Protection.}
The fundamental concept behind obfuscation-based methods is to perturb users' data or model parameter updates, thereby rendering it challenging for adversaries to easily infer sensitive information. 
Broadly speaking, obfuscation-based protection can be categorized into two types: data obfuscation protection and model obfuscation protection.
\begin{itemize}
  \item \textbf{Data Obfuscation.}
  As mentioned before, Wu et al.~\cite{wu2021fedgnn} have identified that user-interacted items can be detected based on non-zero gradients in recommendations from explicit feedback.
  Therefore, Lin et al.~\cite{lin2020fedrec} have devised a hybrid filling strategy to generate synthetic ratings for select items. 
  However, employing such methods excessively can significantly decrease model performance.
FedRS++~\cite{liang2021fedrec++} addresses this issue by utilizing a denoising client group to summarize noisy gradients, assisting the central server in mitigating side effects from fake items.
FeSoG~\cite{10.1145/3501815} employs pseudo-labeling techniques with item sampling for privacy protection.
  Additionally, some works~\cite{10.1145/3460231.3478855,10.1145/3548456} use fake marks to enhance privacy, generating fake gradients for unrated items and fabricating fake ratings.
  
  \item \textbf{Model Obfuscation.} 
  Model obfuscation refers to the method that adds noise to the model gradients or model parameters. Common methods for model obfuscation are the use of Differential Privacy (DP) and Local Differential Privacy (LDP)~\cite{dwork2006calibrating} when sharing model parameters. For example, F2MF \cite{liu2022fairness} proposes to use DP techniques to protect users' group information, thereby achieving the fairness objective in the federated recommendation. 
  PrivateRec \cite{liu2023privaterec} introduces a differentially private attention module to provide a differential privacy (DP) guarantee for uploaded decomposed user embeddings.
    DP-PrivRec \cite{10.1007/s00778-021-00700-6} utilizes Gaussian mechanism \cite{dwork2008differential} to perturb the uploaded gradients and introduces an adaptive gradient clipping mechanism for controlling the contribution of each client. ReFRS \cite{10.1145/3560486} proposes only to upload the parameters of the encoder to the server, and additional DP and homomorphic encryption (HE) techniques are integrated with the parameters for privacy concerns. FedCTR \cite{10.1145/3506715}, in the context of multi-platform federated learning, employs LDP and DP techniques to encrypt the local user embeddings on each platform and the aggregated user embeddings on the server side, respectively. 
    Both ~\cite{wu2021fedgnn} and~\cite{qi2020privacy} implement Laplace Noise-based LDP in FedRSs by clipping gradients and introducing Laplace noise.
    Liu et al.~\cite{10.1145/3501815} explore the addition of a dynamic scale of noise to different gradients based on their magnitudes.
    Zhang et al.~\cite{zhang2023comprehensive} highlight that different model components exhibit varying vulnerabilities to inference attacks. In response, they propose an adaptive mechanism to establish a dynamic privacy budget for each model component. 

    Mask matrix is also widely used in DeviceRss. It obfuscates gradients or parameters by multiplying them with a random matrix.
    For example, VFUCB \cite{cao2023privacy} introduces an orthogonal matrix-based mask mechanism (O3M), which masks the shared user feature data by multiplying with an orthogonal matrix under the cross-platform federated recommendation scenarios. Analogously, FedNCF \cite{10.1016/j.knosys.2022.108441} generates random matrices based on random seeds shared between clients, and these matrices are used as masks by multiplying them with the uploaded parameters. FedeRank \cite{anelli2021federank} allows users to freely control the proportion of the gradient they upload, with the remaining portion being masked as zero.
\end{itemize}

To sum up, whether employing data obfuscation or model obfuscation, additional noise is introduced during the training process. 
Consequently, all of these methods must strike a balance between system utility and privacy protection.

\textbf{Encryption-based Protection.}
Another way to protect user privacy is to use advanced encryption algorithms to encrypt parameters before uploading them to the central server.
In general, there are three commonly used encryption schemes: homomorphic encryption~\cite{gentry2009fully}, secret sharing~\cite{shamir1979share}, and secure multiparty computation (MPC) \cite{chen2020secure}.
\begin{itemize}
  \item \textbf{Homomorphic Encryption.}
  Homomorphic encryption is a form of encryption that enables computational operations on encrypted data without the need for decryption~\cite{gentry2009fully}.
  In DeviceRss, this kind of encryption technique is used to encrypt public parameter gradients~\cite{chai2020secure} or private parameters~\cite{wu2021fedgnn,10.1145/3511808.3557668,luo2023perfedrec++,10.1145/3460231.3478855,perifanis2023fedpoirec}.
  In general, homomorphic encryption can protect users' privacy while maintaining system utility. 
  However, it has some limitations. 
  The first limitation pertains to ensuring the safety of secret keys. 
  The second drawback is that encryption algorithms typically increase computational burden; therefore, these encryption methods may not be applicable to certain edge devices.
  \item \textbf{Secret Sharing.}
  Secret sharing~\cite{shamir1979share} breaks a ``secret'' into multiple pieces and disperses these pieces to multiple participants so that the secret can only be reconstructed when all pieces are gathered.
  Li et al.~\cite{li2016algorithm} employ secret sharing to compute the similarity between every pair of rated items, thus safeguarding the privacy of users' rating values.  Ying et al.~\cite{ying2020shared} apply secret sharing to protect both item ratings and user interaction behavior.
  FR-FMSS~\cite{10.1145/3460231.3478855} and FMSS~\cite{10.1145/3548456} combine both fake items and secret sharing to provide more comprehensive privacy protection.
  Compared to obfuscation-based methods and homomorphic encryption, secret sharing can provide privacy protection with lower model performance compromises and lower computation power; however, the distribution of secret pieces will result in huge communication costs.
    \item \textbf{Secure Multiparty Computation (MPC) \cite{chen2020secure}:} MPC allows multiple clients/platforms to jointly compute a function over their inputs while keeping those inputs private. For example, S$^{3}$Rec \cite{cui2021exploiting} employs MPC and private information retrieval \cite{angel2018pir} techniques to ensure the secure matrix multiplication protocol for the cross-platform federated social recommendation.

\end{itemize}

\subsubsection{Privacy Issues and Solutions from Service Provider Perspectives}
Most current privacy-related works in DeviceRSs mainly investigate how to fulfill users' privacy needs as we discussed in Section~\ref{sec_user_privacy}.
However, as illustrated in Figure~\ref{fig_privacy_perspectives}, service providers also have privacy concerns when using DeviceRSs.
Specifically, in DeviceRSs, recommender models are exposed to all participants as they are deployed on customers' local devices.
Since recommender models are the core intellectual property and the service providers develop them with substantial expenses, especially in commercial scenarios, the guarantee of model information protection is crucial to convince service providers to utilize DeviceRSs.
Besides, how to protect the model IP-right after deploying DeviceRSs is also important for service providers. PTF-FedRec~\cite{yuan2023hide} is a model parameter-transmission-free FedRS framework, allowing the clients and the central server to collaboratively learn heterogeneous models. Based on their framework, the service provider can assign some naive models and algorithms on the client side while deploying the powerful model on the server side. Overall, the protection of model information and model IP-right are still under-explored in DeviceRss.

\subsection{Poisoning Attacks and Countermeasures}\label{section:robustnessrisksandcountermeasures}
\begin{figure}
\setlength{\abovecaptionskip}{1pt}
  \centering
  \subfloat[Data poisoning attack for CloudRSs.]{\includegraphics[width=0.22\textwidth]{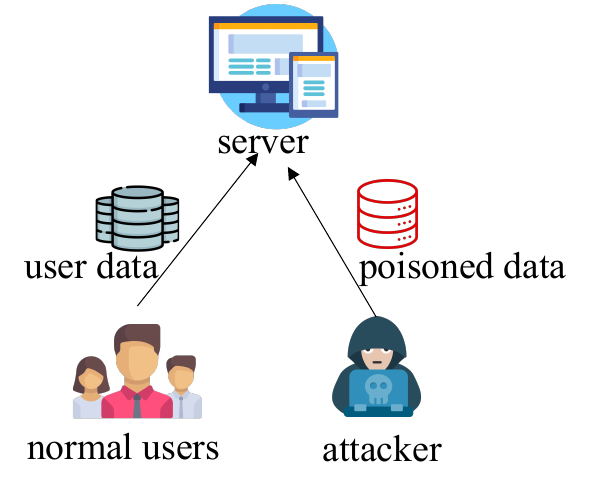}\label{fig_data_cloud}}
  \hfil
  \subfloat[Data poisoning attack for DeviceRSs.]{\includegraphics[width=0.3\textwidth]{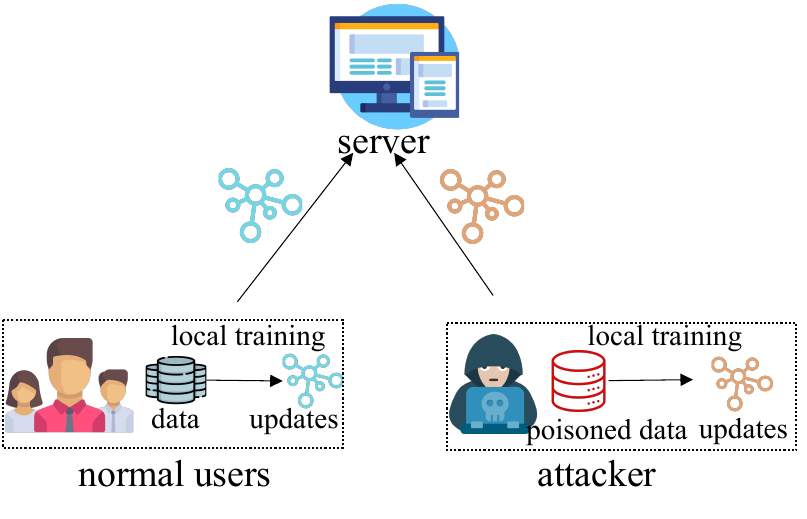}\label{fig_data_device}}
  \hfil
  \subfloat[Model poisoning attack for DeviceRSs.]{\includegraphics[width=0.3\textwidth]{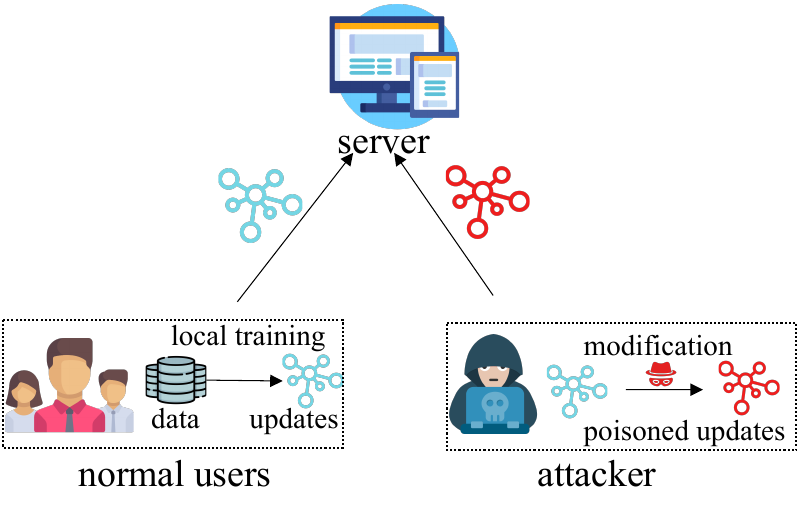}\label{fig_model_device}}
  \caption{Overview of different attacks for CloudRSs and DeviceRSs.}
  \label{fig_attack_types}
\end{figure}

The robustness problem mainly exists in the training process, where adversaries can influence the model using pollution.
In the traditional CloudRSs, the threats for robustness mainly exist in the data integrating stage, where adversaries can manipulate the recommender system to achieve specific malicious goals (e.g., promoting or demonting specific items with financial incentives~\cite{di2020taamr}, degrading the performance of the system~\cite{yu2023untargeted}, etc.) by uploading fake item interactions into the training dataset~\cite{wu2022poisoning,wu2021triple,lin2020attacking,fang2018poisoning,christakopoulou2019adversarial}.
This kind of attack is called a data poisoning attack.

In DeviceRSs, as the users can directly upload model parameters to update the recommender model, the threats for robustness are more complicated.
Specifically, the adversary can not only craft fake interactions to disturb the training process, but also directly upload poisoned model updates to manipulate the system, namely model poisoning attacks.
In the following, we introduce the research of both data poisoning attacks and model poisoning attacks and their corresponding solutions.
It is worth mentioning that the current attack and defense works of DeviceRSs mainly focus on federated recommender systems.
The robustness of other kinds of DeviceRSs, such as semi-decentralized DeviceRSs are still under-explored.
Figure~\ref{fig_attack_types} compares different types of attacks in CloudRSs and DeviceRSs.

\subsubsection{Data Poisoning Attacks and Defense}
Research on data poisoning attacks in the context of DeviceRSs is considerably less than that on model poisoning attacks.
This phenomenon can be attributed to several reasons.
Firstly, traditional data poisoning attacks, such as Bandwagon attacks~\cite{kapoor2017review,gunes2014shilling}, can only be executed under the assumption that a portion of user private data is accessible.
However, in DeviceRSs, launching these data poisoning attacks is more costly than in centralized recommender systems, given that users' data are individually managed by devices. 
To acquire these data, adversaries would need to compromise a large number of devices.
Besides, some studies suggest that these data poisoning attacks are more detectable in DeviceRSs by analyzing the gradients uploaded by clients.
For example, Jiang et al.~\cite{jiang2020detection} design a set of gradient features, encompassing gradient deviation from average agreement, weighted gradient degree, gradient similarity with neighbors, and Filler mean difference.
Then, based on these features, a data poisoning attacker (e.g., HySAD~\cite{wu2012hysad}) can be detected using clustering methods.
FedAttack~\cite{wu2022fedattack} stands out as the sole pure data poisoning attack specifically designed for a typical kind of DeviceRSs (FedRSs) so far.
The goal of FedAttack is to disrupt a federated recommender system with limited Byzantine clients. 
It leverages the observation that globally hardest negative samples are usually detrimental to the system as they have a high chance of being false positive items and guiding the model to converge to local minima ~\cite{xuan2020hard}.

\subsubsection{Model Poisoning Attacks and Defense}
Model poisoning attacks are a specific category of attacks designed for DeviceRSs. 
Exploiting the characteristics of DeviceRSs, which allows all participants to access the recommendation model, adversaries in model poisoning attacks craft poisoned model updates to directly modify the model.

PipAttack~\cite{zhang2022pipattack} stands as the initial exploration into model poisoning attacks in FedRSs. 
It achieves item promotion by aligning the embeddings of target items with popular items, requiring prior knowledge of these popular items.
A drawback of PipAttack is its reliance on controlling a substantial number of malicious clients to ensure effectiveness. To reduce the number of adversarial clients, FedRSAttack~\cite{rong2022fedrecattack} permits attackers to acquire a portion of benign clients' interaction data, using this data to mimic benign user embeddings—an assumption with potential limitations. 
Both PipAttack and FedRSAttack rely on specific prior knowledge beyond the generic learning protocol, limiting their applicability in various scenarios. In response to these limitations, Rong et al.~\cite{rong2022poisoning} propose two model poisoning attacks, A-ra and A-hum, which do not depend on any prior knowledge.
A-ra employs random vectors sampled from a Gaussian distribution to simulate the feature vectors of normal clients, while A-hum further optimizes these random vectors by treating the target items as negative samples to create ``hard users''.
PSMU~\cite{yuan2023manipulating} suggests that incorporating alternative items can enhance the competitiveness of target items.
Yuan et al.~\cite{yuan2023manipulating1} take the pioneering step of combining data poisoning attacks and model poisoning attacks in visually-aware FedRSs.
They promote target items by contaminating both their visual signals (e.g., item posters) and item embeddings, highlighting the potential threats of incorporating third-party images in FedRSs.
In contrast to the aforementioned model poisoning attacks, whose adversarial objective is to manipulate the rank of specific items, Yu et al.~\cite{yu2023untargeted} aim to diminish overall model performance by uploading gradients to guide item embeddings to converge into several dense clusters.

All of the above model poisoning attacks, especially those with more realistic setting~\cite{rong2022poisoning,yuan2023manipulating,yuan2023manipulating1}, consistently underscore the vulnerability of DeviceRSs to model poisoning attacks, attributed to the exposure of the training process. 
To address this security loophole, numerous defense methods have been researched recently.

Some works~\cite{zhang2022pipattack,yuan2023manipulating,yu2023untargeted} have explored defense methods proposed in federated learning.
For instance, Zhang et al.~\cite{zhang2022pipattack} applies Bulyan~\cite{guerraoui2018hidden} and Trimmed Mean~\cite{yin2018byzantine} to defend against their proposed PipAttack.
The experimental results showcase that these two defenses cannot achieve robust performance.
PSMU~\cite{yuan2023manipulating} further employs an item-level Krum~\cite{blanchard2017machine}, Median~\cite{yin2018byzantine}, and gradient clipping~\cite{guerraoui2018hidden} to safeguard recommender systems.
Nevertheless, their research suggests that Krum, Median, and Trimmed Mean significantly compromise model performance.
Yu et al.~\cite{yu2023untargeted} delve into the reasons behind the ineffectiveness of these federated learning defense mechanisms in DeviceRSs. 
The primary challenge lies in the non-IID (non-identically distributed) nature of data for each client, leading to significant gradient deviations for both benign and malicious clients. 
Consequently, distinguishing poisoning gradients in DeviceRSs becomes a formidable task.
The work~\cite{chen2020robust} is one of the pioneers studying the defense method to improve the robustness of DeviceRSs to resist model poisoning attacks.
It proposes A-FRS and A-RFRS to improve Byzantine resilience, however, these defenses are only shown to protect models from gradient ascent attack~\cite{blanchard2017machine}.
HiCS~\cite{yuan2023manipulating} introduces a sparsification update with gradient clipping defense methods which can defend against most model poisoning attacks.
However, this method may slow down the convergence of models. 
Yu et al.~\cite{yu2023untargeted} design a defense named UNION to improve the resilience of FedRSs for untargeted model poisoning attacks.

\section{Evaluation Metrics}\label{sec:metrics}

The evaluation of on-device recommender systems (DeviceRSs) encompasses multiple dimensions beyond standard accuracy. Due to the resource-constrained nature of edge devices and the increasing concern over user privacy, performance metrics must comprehensively reflect accuracy, efficiency, communication cost, and robustness. In this section, we organize the evaluation metrics into four categories: general recommendation accuracy, on-device inference efficiency, training and update cost, and robustness against security and privacy threats.

\subsection{General Recommendation Accuracy}

Accuracy remains the most fundamental measure for recommender systems, with tasks generally falling into three categories: ranking (e.g., top-$K$ recommendation), regression (e.g., rating prediction), and classification (e.g., click-through rate prediction).

\begin{itemize}
    \item \textbf{Hit Ratio@K (HR@K)}: Measures whether the ground-truth item appears in the top-$K$ ranked list. It reflects the system’s recall capability at a fixed cutoff.
    \begin{equation}
        HR@K = \frac{1}{|U|} \sum_{u \in U} \mathbb{I}(r_u \in R_u^K),
    \end{equation}
    where $U$ is the set of users, $r_u$ is the ground-truth item for user $u$, and $R_u^K$ is the top-$K$ recommendations.

    \item \textbf{Recall@K}: Captures the proportion of relevant items successfully retrieved in the top-$K$ list. It is especially relevant when users may have multiple relevant items.
    \begin{equation}
        Recall@K = \frac{1}{|U|} \sum_{u \in U} \frac{|R_u^K \cap T_u|}{|T_u|},
    \end{equation}
    where $T_u$ is the ground-truth item set for user $u$.

    \item \textbf{Normalized Discounted Cumulative Gain (NDCG@K)}: Takes into account the position of relevant items in the top-$K$ list, assigning higher weight to items ranked higher.
    \begin{equation}
        NDCG@K = \frac{1}{|U|} \sum_{u \in U} \frac{DCG_u@K}{IDCG_u@K}, \quad DCG_u@K = \sum_{i=1}^{K} \frac{rel_i}{\log_2(i + 1)},
    \end{equation}
    where $rel_i$ denotes the binary relevance of the item at position $i$, and $IDCG_u@K$ is the ideal DCG score.

    \item \textbf{Mean Squared Error (MSE)}: Commonly used in rating prediction tasks, MSE measures the average squared difference between predicted scores and ground-truth ratings.
    \begin{equation}
        MSE = \frac{1}{|D|} \sum_{(u, i) \in D} (\hat{r}_{ui} - r_{ui})^2,
    \end{equation}
    where $D$ is the set of user-item pairs, and $\hat{r}_{ui}$ and $r_{ui}$ denote the predicted and actual ratings respectively.

    \item \textbf{AUC (Area Under the ROC Curve)}: Used primarily in binary classification tasks such as click-through rate (CTR) prediction. AUC reflects the probability that a randomly selected positive instance is ranked above a negative one.
\end{itemize}

\subsection{On-Device Deployment and Inference Efficiency}

DeviceRSs are designed to operate under strict resource constraints. Therefore, evaluation of model footprint and real-time response efficiency becomes critical.

\begin{itemize}
    \item \textbf{Model Size}: Measures the storage footprint of the model, often expressed in megabytes (MB) or number of parameters. Smaller models are better suited for mobile and embedded environments with limited storage and memory.

    \item \textbf{Inference Time}: The average time required to produce recommendations for a single user or session on-device. Typically measured in milliseconds, it directly impacts user experience, especially for interactive applications.
\end{itemize}

\subsection{Training and Update Efficiency}

On-device recommender systems often rely on federated or decentralized training frameworks to avoid transmitting raw user data. In this context, communication and convergence efficiency are essential.

\begin{itemize}
    \item \textbf{Communication Cost}: Refers to the volume of data transferred between client devices and the central server during each training round. It includes model updates, gradients, or compressed representations, and is typically measured in megabytes per round.

    \item \textbf{Convergence Time}: Indicates the number of training rounds or total elapsed time required to reach a performance plateau. Faster convergence is preferred for frequent model updates in dynamic environments.
\end{itemize}

\subsection{Security and Privacy Robustness}

In federated and on-device settings, user data is kept local, but model updates may still leak private information. Robustness against inference attacks and adversarial manipulations is thus a critical evaluation dimension.

\begin{itemize}
    \item \textbf{F1 Score (Attribute Inference Attack Resistance)}: Measures the ability of an attacker to infer sensitive user attributes (e.g., gender, age) from exposed model updates or embeddings. A lower F1 score from the attacker’s classifier implies higher privacy protection by the system. This metric is especially relevant in federated learning scenarios.

    \item \textbf{Exposure Rate (ER@K)}: Designed to evaluate a system's vulnerability to poisoning or promotion attacks. In such attacks, an adversary aims to manipulate the model so that a specific \emph{target item} appears more frequently in user recommendation lists. ER@K quantifies how often the target item is exposed among the top-$K$ ranked items:
    \begin{equation}
        ER@K = \frac{1}{|U|} \sum_{u \in U} \mathbb{I}(t \in R_u^K),
    \end{equation}
    where $t$ is the adversarial target item, and $R_u^K$ is the top-$K$ recommendation list for user $u$. A high ER@K score suggests that the model is more susceptible to targeted exposure, indicating lower robustness.
\end{itemize}

\section{Challenges and Further Directions}\label{section:challengesanddirections}
Although DeviceRSs have shown promising performance,  we will discuss the challenges they still face and potential future research directions.
\subsection{Heterogeneity in DeviceRSs}
Most DeviceRS methods assume that each device/user is homogeneous, but this assumption is difficult to satisfy in real life due to the inherent heterogeneity among devices/users, which mainly includes data heterogeneity, system heterogeneity, and privacy heterogeneity. Data heterogeneity refers to the distribution of data on each client side, which is heterogeneous, usually non-independent and identically distributed \cite{desai2023heterogeneous}. System heterogeneity refers to each device's performance, e.g., storage, computation, and communication capabilities are different. Overlooking the heterogeneity will inevitably degrade the performance of the model \cite{yuan2023hetefedrec}. Privacy heterogeneity refers to users could have a different privacy budget, with some users not minding sharing their private data with platforms in exchanges for better recommendation service \cite{anelli2021federank,qu2024towards}.

\subsection{Fairness in DeviceRSs}
On top of the heterogeneity challenge mentioned above, another future challenge that needs to be focused on is the fairness of DeviceRSs \cite{maeng2022towards}. For example, in federated recommendation methods, when utilizing FedAvg, clients with larger datasets are typically assigned larger aggregation weights. This leads to a bias where recommendations are skewed in favor of those clients with more data. As a result, the system may not accurately represent or serve the interests of clients with smaller datasets, potentially impacting the overall effectiveness and fairness of the recommendation system. 

\subsection{Evolving User Dynamics in DeviceRSs}
Most current DeviceRSs assume that the number of users remains constant, which is often not the case in real-life scenarios. To address this, future research should focus on two main areas: (1) Cold start for DeviceRSs: the cold-start problem in DeviceRSs presents a more complex challenge compared to traditional CloudRSs. While CloudRSs focus on generating recommendations for new users or items with minimal historical data, DeviceRSs face a new additional task of deploying models to newly added devices. This dual challenge involves not only providing accurate recommendations for cold-start users or items but also ensuring rapid model deployment, coupled with the requirement to cater to varying device capabilities.
(2) Unlearning for DeviceRSs: When users exit the system, it is essential to handle their data appropriately. This is where ``federated unlearning \cite{yuan2023federated}'' comes into play, a process of selectively forgetting or disregarding the data of users who are no longer active. This approach not only maintains the accuracy and relevance of the system's recommendations but also addresses privacy concerns.

\subsection{Model Copyright Protection in DeviceRSs}
Unlike CloudRSs, DeviceRSs expose recommender models to all users. In practical scenarios, especially in commercial applications, recommender models represent core intellectual properties (IP) that entail significant efforts and financial investments from various experts. Therefore, it is crucial to explore methods for protecting Model IP in DeviceRSs settings. This protection can be approached from two aspects: (1) preventing the model copy behavior and (2) detecting illegal model copy behavior.~\cite{yuan2023hide} delves into preventing model copy behavior by achieving heterogeneity in clients' and the server's models. Additionally, encryption methods can be considered to thwart illegal model extraction. To detect illegal model copy behavior, one potential approach involves embedding watermarks in recommender models.

\subsection{Foundation Models in DeviceRSs}
Inspired by the remarkable progress of foundational models across various research domains, such as natural language processing~\cite{min2023recent} and computer vision~\cite{du2022survey}, there is growing interest in developing foundational models specifically for recommender systems~\cite{fan2023recommender}. Most existing efforts~\cite{he2023large,wei2023llmrec,cui2022m6,liu2023chatgpt} are based on cloud-centric architectures, which inherently face challenges such as response latency. However, due to their large-scale nature, transferring these foundational models to on-device recommender systems (DeviceRS) is nearly infeasible with current techniques. Future research could explore lightweighting these models and leveraging cloud-edge-device collaboration~\cite{yuan2024fellas,long2024diffusion} to efficiently deploy them, addressing scalability challenges while maintaining performance.

\subsection{Quantitative Evaluation and Benchmarking in DeviceRSs}
Current research on DeviceRSs is conducted using various datasets and evaluated through diverse evaluation metrics (e.g., recall, communication costs, memory consumption), experimental settings (e.g., leave-one-out validation, split-by-ratio for training, validation, and testing), and tasks (e.g., top-k recommendation, sequential recommendation, CTR prediction). However, the absence of a standardized benchmarking framework presents significant challenges in comparing these systems and deriving consistent conclusions about their performance. Therefore, there is an urgent need for a comprehensive evaluation framework that unifies different tasks and experimental settings, providing a more comprehensive and fair comparison across systems. Such a framework would facilitate the standardization of experimental environments and metrics, offering a clearer understanding of the trade-offs involved in deploying recommender systems on resource-constrained devices.

\section{Conclusion}\label{section:conclusion}
The migration of cloud-based recommender systems (CloudRSs) to device-based recommender systems (DeviceRSs) has become a research trend in both academia and industry. In this survey, we comprehensively reviewed research on DeviceRSs from three aspects: model deployment and inference, training and updating, as well as security and privacy. For each aspect, we systematically categorized the existing methods, summarized the main architectures, and outlined the strategies adopted by different methods. Additionally, we discussed the current challenges in this research area and introduced potential future research directions. We hope that this review will provide readers with a thorough understanding of this field.

\bibliography{02-main}

\end{document}